\documentclass[useAMS,usenatbib]{mn2e}

\DeclareTextSymbol{\degre}{OT1}{23}
\usepackage{graphicx}
\usepackage{amsmath}
\usepackage{multirow}
\title[Accurate SB2 orbits and masses]{Masses of the components of SB2 binaries observed with {\it Gaia}. V. 
Accurate SB2 orbits for 10 binaries and masses of the components of 5 binaries.
\footnotemark[1]\thanks{based on observations performed
at the Observatoire de Haute--Provence (CNRS), France}
\footnotemark[2]\thanks{based on data obtained with the ESO Very Large Telescope under
programme 094.D-0624 and 097.D-0688.}
}
\author[J.-L. Halbwachs et al.]{J.-L. Halbwachs$^{1}$\thanks{E-mail: jean-louis.halbwachs@astro.unistra.fr},
F. Kiefer$^{2,3}$, Y. Lebreton$^{3,4}$, H.M.J. Boffin$^{5}$,
F. Arenou$^{6}$, 
\newauthor J.-B. Le Bouquin$^{7}$, B. Famaey$^{1}$, D. Pourbaix$^{8}$, P. Guillout$^{1}$, J.-B. Salomon$^{9}$,
\newauthor and T. Mazeh$^{10}$ \\
$^{1}$Universit\'e de Strasbourg, CNRS, Observatoire astronomique de Strasbourg, UMR 7550, 
11 rue de l'Universit\'{e}, F-67000 Strasbourg, France\\
$^{2}$Institut d'Astrophysique de Paris,CNRS/UMR7095, 98bis boulevard Arago, F-75014 Paris, France \\
$^{3}$LESIA, Observatoire de Paris, PSL  Research  University, CNRS, Sorbonne Universit\'e, UPMC Univ. Paris 06, Univ. Paris Diderot,\\
Sorbonne Paris Cit\'e, F-92195 Meudon, France\\
$^{4}$Univ Rennes, CNRS, IPR (Institut de Physique de Rennes) - UMR 6251, F-35000 Rennes, France\\
$^{5}$European Southern Observatory, Karl-Schwarzschild-Strasse 2, D-85748 Garching bei M\"unchen, Germany \\
$^{6}$GEPI, Observatoire de Paris, PSL Research University, CNRS, Universit\'e Paris Diderot, Sorbonne Paris Cit\'e, Place Jules Janssen,\\
F-92195 Meudon, France\\
$^{7}$Univ. Grenoble-Alpes, CNRS, IPAG, 38000 Grenoble, France \\
$^{8}$FNRS, Institut d'Astronomie et d'Astrophysique, Universit\'{e} Libre de Bruxelles, boulevard du Triomphe, 1050 Bruxelles, Belgium\\
$^{9}$Institut UTINAM, CNRS UMR6213, Universit\'e Bourgogne Franche-Comt\'e, OSU THETA, Observatoire de Besan\c con, BP 1615, F-25010\\
Besan\c con Cedex, France\\
$^{10}$School of Physics and Astronomy, Tel Aviv University, Tel Aviv 69978, Israel\\
}

\begin{document}

\date{Accepted . Received 2019 ; in original form 2019}

\pagerange{\pageref{firstpage}--\pageref{lastpage}} \pubyear{2019}

\maketitle

\label{firstpage}

\begin{abstract} 
Double-lined spectroscopic binaries (SB2s) are one of the main sources of stellar masses,
as additional observations are only needed to give the inclinations of the orbital planes
in order to
obtain the individual masses of the components. For this reason, we are observing a selection
of SB2s using the SOPHIE spectrograph at the Haute-Provence observatory in order to precisely
determine their orbital elements. Our objective is to finally obtain masses with an accuracy 
of the order of one percent by combining our radial velocity (RV) measurements and the astrometric
measurements that will come from the {\it Gaia} satellite. We present here the RVs and the
re-determined orbits of 10 SB2s. In order to verify the masses we will derive from {\it Gaia}, 
we obtained interferometric measurements of the ESO VLTI for one of these SB2s. Adding the
interferometric or speckle measurements already published by us or by others for 4 other stars, we 
finally obtain the masses of the components of 5 binary stars, with masses ranging from 0.51 to 2.2 
solar masses, including main-sequence dwarfs and some more evolved stars whose location in the HR diagram
has been estimated.
\end{abstract}

\begin{keywords}
binaries: spectroscopic, stars: fundamental parameters, stars: individual: HIP 104987
\end{keywords}

%=======================================================================================================================

\section{Introduction}

Estimating the mass of stars is a fundamental step in understanding the internal processes that determine how
stars work, how bright they shine, and for how long. 
When the mass is known with sufficient precision, modelling not only enlightens us on the physical processes it depicts
\citep[e.g.][]{Claret19},
but also provides constraints on parameters that are not directly accessible, such as age and helium content 
\citep[e.g.][]{Lebreton05}. 

It is only possible to obtain a mass for a few very specific stars, but these are the basis of calibration 
relationships that allow the masses of other stars to be estimated from more accessible data, such as spectral 
type, color indices or absolute luminosity \citep[e.g.][]{Eker15,Moya18,Mann19}.
These calibrations paved the way for population synthesis models, ultimately allowing to estimate the stellar mass-to-light
ratios of galaxies from their observed colours or spectral energy distributions \citep[e.g.][]{Bell01}, which are central to
the estimate of their dark matter content \citep[e.g.][]{Lelli16}.

There are only a limited number of direct methods for measuring the mass of a star, and they all apply to 
components of double stars. Most (but not all) use the double-lined spectroscopic binaries (SB2s), for which the orbital 
elements allow to calculate a minimum value of the mass of each component, ${\cal M} \sin^3 i$, where $i$ is the
inclination of the orbit. This parameter can be obtained from the following complementary techniques:

\begin{itemize}

\item
The eclipsing binaries (EBs) have historically been considered the Royal Road of stellar astrophysics, and this reputation
has not been denied. When they are also double-lined spectroscopic binaries (SB2s), photometric and spectroscopic 
observations make
it possible to deduce not only the masses of the components, but also their radii, their effective temperatures
\citep[e.g.][]{Torres19}, and even the distance of the system with sufficient precision to test trigonometric parallaxes
obtained by astrometric satellites \citep{Munari04}. 
Unfortunately, a binary can present eclipses only if its orbital plane is close to the line of sight. This makes the EBs
quite rare among binaries, and introduces a bias in favour of short-period systems. As a result, the components of many 
EBs are affected by the presence of the companion, and only a minority of EBs are representative of single stars.

\item
Visual binaries (VBs) are another case where stellar masses can be obtained, when they are also SB2s.
Formerly confined to the domain of long periods, they have moved into the domain of moderate periods (from a few weeks to
decades) thanks to the development of interferometry, whether speckle \citep[e.g.][]{McAlister96, Balega07} or long-baseline
\citep[see e.g.][]{Pionier11}. The results are even better for EB, VB and SB2 systems combined
\citep[e.g.][]{Lester19, Gallenne19}.
However, the number of short period VB systems is still limited by the considerable resources required 
to observe an orbit, by the brightness required to observe a star and by the low luminosity contrast between 
the two components.
 
\item
Astrometric binaries (ABs). We call here ``astrometric binaries'' unresolved double stars whose photocentre describes 
a measurable orbit, such as the 235 orbits observed by the ``HIgh Precision PARallax COllecting Satellite''
\citep[Hipparcos,][]{ESA97}. The masses of the AB components may be derived when the system is also a VB
\citep[][]{Martin98}, or an SB2 \citep[][]{Jancart05}. The masses thus obtained have an accuracy of a few percent, 
at the best. However, much more precise masses should be obtained in the near future, thanks to astrometric 
measurements from the Gaia satellite.

\end{itemize}

Compilations of the most accurate masses has been given by \cite{Torres10}, and in the references mentioned above 
about the calibrations.

This paper is the fifth in a series dedicated to the determination of precise masses using the {\it Gaia} satellite 
\citep{Gaia16}.
Although the Gaia collaboration has already published two Data Releases \citep[][]{GaiaDR1,GaiaDR2}, precise masses can only be 
calculated when the full data transits are available, with the full release for the nominal mission, according to the Gaia web
site\footnote{https://www.cosmos.esa.int/web/gaia/release}.

The first paper of the series \citep[][Paper I hereafter]{Halb2014} presented the selection of about 70 SB2s for which the masses 
of the components could be precisely calculated by combining the astrometric transits of the {\it Gaia} satellite with precise
radial velocity (RV) measurements obtained using the ``Spectrographe pour l'Observation des PH\'enom\`enes des Int\'erieurs Stellaires et des Exoplan\`etes''
\citep[SOPHIE,][]{Perruchot} at the Haute-Provence Observatory. The selection
included about fifty known SB2s, and about twenty SB1s that our first spectroscopic observations had transformed into SB2s by
detecting the secondary component.

Simultaneously with the spectroscopic observations, we obtained interferometric observations for five stars 
of our selection. These 
observations were carried out with the auxiliary telescopes of the ESO Very Large Telescope with the 
``Precision Integrated-Optics Near-infrared Imaging ExpeRiment'' (PIONIER) instrument. In the second
paper \citep[][Paper II hereafter]{Halb2015}, they were already used to derive preliminary masses for the components of two SB2s, using 
published RV measurements completed with a few ones that we had obtained from a preliminary reduction of our {\sc sophie} observations. 
We have thus demonstrated the possibility of using these measurements to validate the masses
that we will later obtain from {\it Gaia}.

In the third and in the fourth paper, \citep[][Paper III and Paper IV, respectively]{Kiefer2016,Kiefer2018}, we presented the RV measurements and
the revised SB2 orbits of 10 and 14 binaries, respectively. 
Out of a total of 24 SB2s revised orbits, we found four for which an interferometric orbit had already been published.
By combining the interferometric measurements of these stars with our RV measurements, we have calculated the masses of the components of these
four binaries (one in Paper III and three in Paper IV).

The present paper is particularly in line with the latter two, since we apply
the same methods to treat 10 more SB2s. Five of these binary stars were resolved by long-base or speckle interferomery: 
One was found in the Fourth Catalog of Interferometric Measurements of Binary 
Stars\footnote{https://www.usno.navy.mil/USNO/astrometry/optical-IR-prod/wds/int4} 
\citep[INT4 hereafter; Third catalogue:][]{Hartkopf01}, one was observed by \cite{Balega07},
and three have been observed for us with the PIONIER instrument attached to ESO's
Very Large Telescope Interferometer.
The interferometric orbits of two of the latter were derived in Paper II, but the third is calculated here for the first time.
Thus, we give here the masses of the components of the five binaries.

The article is organized as follows: the observations are presented in Section~\ref{sect:observations}; this section includes the spectroscopic
observations of 10 SB2s, but also the interferometric observations of one of these stars. 
The derivation of the RVs is in Section~\ref{sect:RV}. The elements of the spectroscopic orbits are derived in Section~\ref{sect:SB2orbits}. The masses of
the components of 5 binaries separated by interferometry are derived in Section~\ref{sect:interfero}, where we briefly
discuss the evolutionary state of these stars and their positions in the HR diagram. Section~\ref{sect:conclusion} is the
conclusion.

%===========================================================================================================================================

\section{Observations}
\label{sect:observations}

\subsection{Spectroscopic observations}
\label{sect:observationsSpectro}

As before, the observations were performed at the T193 telescope of the Haute-Provence Observatory, with the SOPHIE spectrograph.
From 2010 to 2016, they were carried out in visitor mode by assigning priorities to the stars based on ephemerides. Since
semester 2014B, we have regularly obtained observations in service mode that we request for selected dates in order to complete 
phase coverage while avoiding blends that would give unusable RV measurements. 
Exceptions to this rule are generally stars for which the preliminary orbit was inaccurate, or stars suspected for a time 
of being multiple systems.

The list of stars treated in this article is given in Table~\ref{tab:obs}, where the number of usable spectra and 
the number of cycles covered by the observations are indicated. A minimum of 11 usable spectra was requested in order to calculate
orbital elements with reliable uncertainties.

The exposure times have been adapted to the observation conditions in order to have a signal-to-noise ratio (SNR) appropriate for
each star. The SNR is a compromise between the need to have sufficiently smooth spectra to distinguish the components, and the 
need to have exposure times shorter than one hour, which is the limit in service mode.

\begin{table}
\caption{The SB2s analyzed in this paper.}
\small
\begin{tabular}{@{}l@{~~}l@{~~}c@{~~}c@{~~}c@{~~}c@{~~}l@{}}
\hline
Name       & Alt. name  &  V       & Period$^a$ & $N_\text{spec}$ $^b$ & Span$^c$ & SNR$^d$  \\
HIP        &  HD/BD     &(mag.)    &  (day)     &                      &(period)  &          \\
\hline    
\multicolumn{7}{c}{\it Previously published SB2} \\
&&&&&& \\
HIP\,20601      & HD\,27935     & 8.93    & 156       & 16              & 14           & 50   \\
HIP\,73449      & HD\,132756    & 7.31    & 2529      & 11              &  0.88        & 97   \\
HIP\,76006      & HD\,138525    & 6.39    & 582       & 12              &  4.6         & 142  \\
HIP\,77725      & BD\,+11\,2874 & 9.36    & 1016      & 13              &  1.9         & 53   \\
HIP\,96656      & HD\,186922    & 8.04    & 4347      & 14              &  1.0         & 102  \\
HIP\,104987     & HD\,202447/8  & 3.93    & 99        & 14              & 12           & 371  \\
HIP\,117186     & HD\,222995    & 7.11    & 86        & 14              & 26           & 96   \\
\hline
\multicolumn{7}{c}{\it SB2s identified in Paper I, previously published as SB1s} \\
&&&&&& \\
HIP\,7134       & HD\,9313      & 7.81    &  53.5     & 16              & 50           & 98    \\
HIP\,61732      & BD\,+17\,2512 & 9.18    & 595       & 11              &  4.3         & 46   \\
HIP\,101452     & HD\,196133    & 6.70    &  88       & 11              & 30           &129   \\
\hline
\end{tabular}
\flushleft
$^a$ The period values are taken from our solutions. \\
$^b$ $ N_\text{spec}$ gives the number of spectra collected with SOPHIE and taken into account in the derivation of the orbital elements. \\
$^c$ Span is the total time span of the observation epochs used in the orbit derivation, counted in number of periods. \\
$^d$ SNR is the median signal-to-noise ratio of all the SOPHIE spectra of a given star at 5550\,\AA.
\label{tab:obs}
%}
\end{table}

The spectra are used to derive the RVs of the components, as explained in Section~\ref{sect:RV} hereafter.

%---------------------------------------------

\subsection{Interferometric measurements}
\label{sect:interferObservations}

We obtained additional interferometric observations for one of the 10 SB2s, namely HIP 104987. This star was observed with the four 1.8 m
Auxiliary Telescopes of ESO VLTI, using the PIONIER instrument  \citep{Berger10, Pionier11} in the H-band. Twelve set of observations were made, all
resulting in the separation of the components. A first set of observations were done in Visitor Mode under Prog. ID 094.D-0624(A-F) on the nights 
of 6, 8, 17, 18, and 31 October 2014 for a total of 106 data points. The baseline was A1-G1-K0-J3. 
In addition, data were obtained in Service Mode, under Prog. ID 097.D-0688(A), on 7 epochs between 29 May 2016 and 25 August 2018. Each time 2
sets of observations were obtained, leading to a total of 70 data points. The baseline was A0-G1-J2-J3. 

The observations were reduced with the {\sc pndrs} package presented by \cite{Pionier11}. For each epoch, the visibilities and closure phases were
fitted to a binary model to determine the relative separation between the components, $\rho$, the position angle of the secondary component
with respect to the primary, $\theta$, and the flux ratio in the $H$ band. The binary model is non-linear, and $\chi^2$ minimization can lead to
several local minima. Therefore, a classical gridding approach is used to locate the deepest minimum in parameter space. A Levenberg-Marquardt 
algorithm is then used to derive the best-fitting parameters and the covariance matrix, from which are extracted the following parameters of the
astrometric error ellipsoid: the semi-major axis, $\sigma_a$, the semi--minor axis, $\sigma_b$, and the position angle of the semi-major axis,
$\theta_a$; by construction, $\theta_a$ is between 0 and 180$^{\rm o}$. These parameters are presented in Section~\ref{sect:HIP104987}, where they
are used to calculate the visual orbit of HIP~104987, then the masses of the components.

%===========================================================================================================================================

\section{Radial velocity measurements}
\label{sect:RV}

%-------------------------------------------------------------

\subsection{Choice of spectroscopic templates}
\label{sect:templates}

\newcommand{\MS}{\text{\tiny MS}}
\begin{table*}
\centering
\caption{\label{tab:stellpar}The stellar parameters of the 10 SB2s, determined
from the PHOENIX library by $\chi^2$ optimization around the Ca\,I line at 6121\,\AA. 
Sun's parameters derived with the same protocol are given in the last row.}
\scriptsize
\begin{tabular}{@{}lllrrrrr@{}}
\hline %--------------------------------------------------------------
HIP/Name     & $^a$\,\!$ T_\text{eff,1}$ &  $^b$\,\!$ \log g_1$   & $ V_1 \sin i_1$  &  $^c$\,\!$ [\text{Fe/H}]$ &  $\alpha$  & $N_\text{spec}$ & Spectral orders   \\ 
             & $ T_\text{eff,2}$ &  $ \log g_2$   &  $ V_2 \sin i_2$&                  &                & & Median wavelength \\ 
            & (K)               &  (dex)         &   (km s$^{-1}$)   &     (dex)        & (flux ratio)   & & (\AA)\\ 
\hline %--------------------------------------------------------------   

HIP\,7134   & 4754 $\pm$ 48  & 3.75 $\pm$ 0.12        & 4.2 $\pm$ 0.7  & -0.31 $\pm$ 0.04 & 0.045 $\pm$ 0.005 & 4 & 33 \\ 
            & 5057 $\pm$ 243 & 5.13 $\pm$ 0.18	      & 0 (fixed)      &                  &          &	& 6142         \\ 
\\
HIP\,20601  & 5628 $\pm$ 67  & 4.55 $\pm$ 0.07        & 3.4 $\pm$ 1.1  & -0.17 $\pm$ 0.05 & 0.105 $\pm$ 0.004 & 4 & 33  \\ 
            & 4847 $\pm$ 83  & 5.18 $\pm$ 0.10        & 1.1 $\pm$ 0.7  &                  &           &	& 6142                 \\ 
\\
HIP\,61732  & 6021 $\pm$ 16  & 4.44 $\pm$ 0.00$_{MS}$ &  5.8 $\pm$ 0.1 &  0.16 $\pm$ 0.05 & 0.115 $\pm$ 0.009 & 2 & 24, 33   \\  
            & 5070 $\pm$ 326 & 4.59 $\pm$ 0.06$_{MS}$ &  3.0 $\pm$ 0.8 &                  &          &	& 5293, 6142                   \\ 
\\
HIP\,73449  & 5500 $\pm$ 110 & 4.42 $\pm$ 0.08        &  3.8 $\pm$ 0.3 & -0.39 $\pm$ 0.10 & 0.781 $\pm$ 0.115 & 4 & 33   \\ 
            & 5400 $\pm$ 71  & 4.47 $\pm$ 0.07        &  4.7 $\pm$ 0.9 &                  &            &	& 6142                 \\ 
\\

HIP\,76006  & 6314 $\pm$ 26  & 4.10 $\pm$ 0.04        &  8.8 $\pm$ 1.1 & -0.03 $\pm$ 0.04 & 0.157 $\pm$ 0.034 & 4 & 24, 33   \\ 
            & 6083 $\pm$ 152 & 4.72 $\pm$ 0.14        &  4.7 $\pm$ 0.4 &                  &             &	& 5293, 6142                \\ 
\\ 
HIP\,77725  & 4378 $\pm$ 60  & 5.49 $\pm$ 0.05        &  2.7 $\pm$ 0.6 & -0.11 $\pm$ 0.04 & 0.972 $\pm$ 0.051 & 3 & 33   \\ 
            & 4323 $\pm$ 29  & 5.45 $\pm$ 0.07        &  3.2 $\pm$ 0.9 &                  &           &	& 6142                  \\ 
\\
HIP\,96656  & 5128 $\pm$ 8   & 4.58$_{MS}$            &  3.2 $\pm$ 0.5 & -0.37 $\pm$ 0.06 & 0.486 $\pm$ 0.033 & 2 & 33   \\ 
            & 4876 $\pm$ 36  & 4.63$_{MS}$            &  3.8 $\pm$ 0.1 &                  &              &	& 6142               \\ 
\\
HIP\,101452 & 9767 $\pm$ 288 & 3.08 $\pm$ 0.10        & 21.7 $\pm$ 0.1 &  0.08 $\pm$ 0.06 & 0.254 $\pm$ 0.075 & 4 & 24, 33   \\  
            & 7915 $\pm$ 552 & 3.24 $\pm$ 0.13        & 32.2 $\pm$ 2.1 &                  &          &	& 5293, 6142                   \\ 
\\
HIP\,104987 & 5111 $\pm$  7  & 3.08 $\pm$ 0.11        &  5.1 $\pm$ 0.1 & -0.09 $\pm$ 0.01 & 0.814 $\pm$ 0.032 & 4 & 24, 33   \\ 
            & 7488 $\pm$ 223 & 3.87 $\pm$ 0.15        & 23.3 $\pm$ 5.9 &                  &           &	& 5293, 6142          \\

\\
HIP\,117186 & 6208 $\pm$ 138 & 3.05 $\pm$ 0.17        & 42.1 $\pm$ 1.4 & -0.70 $\pm$ 0.02 & 0.347 $\pm$ 0.008 & 4 & 24, 33   \\ 
            & 5785 $\pm$ 110 & 3.27 $\pm$ 0.19        & 13.2 $\pm$ 0.4 &                  &            &	& 5293, 6142         \\ 
\\
Sun         & 5836 $\pm$ 40  & 4.58 $\pm$ 0.10        &  4.9 $\pm$ 0.2 & -0.12 $\pm$ 0.04 &  & 4 & 33                    \\ 
            &  			  & 				      &  			 &				 &  & 4 & 6142                    \\ 
\hline
%--------------------------------------------------------------
\end{tabular}
\flushleft
$^a$Minimum systematic uncertainties on T$_\text{eff}$ are about 100\,K. \\
$^b$The MS subscript indicates that the $\log g$ did not converge to a realistic value ($>5$) and was fixed to be on the Main Sequence following $\log g = 12 - 2\log T_\text{eff}$~\citep{Angelov1996}. \\
$^c$Given the systematic error on [Fe/H]$_\text{sun}$, a more reliable value of uncertainty on [Fe/H] should be at least 0.1\,dex.
\end{table*}

Reliable RV measurements first require the choice of spectroscopic templates. As for the mask used in ordinary 
1D-cross correlation function (1D-CCF), the choice of
templates with a set of absorption lines as similar as possible to the actual absorption lines in the observed spectrum is 
crucial to the estimation of velocities, and to the value of the resulting masses. This is even more important in the present case, when the observed spectrum
is the combination of two different components with different or similar sets of absorption lines. For that reason, the choice of the stellar parameters
that characterise a spectrum was carefully optimised as explained hereafter.

Before any further analysis, the SOPHIE multi-order spectra are reduced, flattened and normalised as explained hereafter:
The spectra are deblazed, flattened, and the pseudo-continuum are normalized using a $p$-percentile filter \citep{Hodg1985}.
The $\chi^2$ of the residuals of the prepared observed spectrum fitted by the sum of two similarly prepared model atmosphere templates 
from the PHOENIX database (Huber et al. 2011) is minimised with respect to $T_\text{eff}$, $\log g$, [Fe/H], $v\sin i$ and flux ratio 
$\alpha=F_B/F_A$ at 4916\,\AA. Order 33 around the Ca\,I line 
at ∼6120\,\AA was mainly used, but early type stars, such as HIP\,61732\,A, HIP\,76006\,A\&B, HIP\,101452\,A\&B, HIP\,104987\,B and HIP117186\,A, 
required the additional use of order 24, bluer and with deep and more numerous lines than on the red wing of their spectrum. When possible and for
each binary, the stellar parameters were optimised for up to four observed spectra with the largest RV separation between the two components of the
binary. In some case, the recursive algorithm leads to unreasonably low or high $\log g$. In those case, main sequence relation of $\log g$ with 
$T_\text{eff}$ is assumed. The results of this preliminary step are presented in Table~\ref{tab:stellpar}.
As in paper IV, this table also gives the results of our method applied to the Sun, using Ceres and Vesta spectra.
Since we measured the Sun metallicity to be of -0.12 dex, this could be considered as the realistic minimum uncertainty on the metallicities that
we derived.

It should be noted that, since the minimisation of the parameters is obtained by a recursive algorithm, the resulting parameters are 
not full-proof against systematics, because metallicity [Fe/H] and effective temperature $T_\text{eff}$ can be degenerate on ranges of order
 $\pm$400\,K and $\pm$0.5\,dex, especially when $v \sin i$ strongly departs from 0. The derivation of the Sun's parameters using
spectra observed with SOPHIE on Ceres and Vesta (see Table~\ref{tab:stellpar}) shows that the effective temperature and surface gravity are
correctly derived within 100\,K and 0.1\,dex but the metallicity is underestimated by 0.12\,dex. 
Nevertheless, the derived model templates are the best-matching with respect to the observed spectra and lead to the best precision possible for RV
derivation of the two binary components, as explained below.

%------------------------------------------------------

\subsection{Derivation of the RVs using {\sc todmor}}
\label{Sect:TODMOR}

After the templates have been fixed, the RVs of the components are derived using {\sc todmor}, which is the multi-order version of the
Two-Dimensional Cross-Correlation algorithm {\sc todcor} \citep{zucker94,zucker04}. 
{\sc todmor} consists in cross-correlating the observed spectra with the sum of two templates each shifted with independent values of
Doppler shift. All orders of the spectra are taken into account, except the few red orders 
with strong telluric absorptions. This leads to a direct measurement of each SB2 component radial velocities from the location of the 2D-CCF peak position.

The RV uncertainties are given by the Hessian of the CCF peak, as explained in \cite{zucker04}. These uncertainties are intrinsic to
the observed spectrum and do not reflect instrument systematics or Earth atmospheric turbulence effects. Therefore, they are generally underestimated.

The resulting RVs and uncertainties are in Table 3. The orbital elements are calculated by correcting these misestimated uncertainties,
as described in Section~\ref{sect:SB2orbits} below.

\begin{table*} 
%\centering
\caption{\label{tab:RVs} New radial velocities from SOPHIE and obtained with {\sc todmor}. The uncertainties must still be corrected
as explained in Section~\ref{sect:SB2orbits}.
Outliers are marked with an asterisk ($^*$) and are not taken into account in the analysis.}
\scriptsize %\small %
\begin{minipage}{89mm}
\begin{tabular}{@{}l@{~~}c@{~~}c@{~~}c@{~~}c@{~~}c@{~~}c@{}}                                                            
\hline                                                                                                                  
\multicolumn{7}{c}{HIP 7134       } \\                                                                                  
&&&&&&   \\                                                                                                             
BJD      & $RV_1$        & $\sigma_{RV 1}$ & $RV_2$        & $\sigma_{RV 2}$ & $O_1-C_1$   & $O_2-C_2$ \\               
-2400000     & km s$^{-1}$   & km s$^{-1}$      & km s$^{-1}$   & km s$^{-1}$      & km s$^{-1}$ & km s$^{-1}$ \\       
\hline                                                                                                                  
55440.5885     &        10.4191     &      0.0065     &        -50.632     &       0.121     &       -0.0194     &        -0.124 \\             
55783.5935$^*$ &       -16.9886$^*$ &      0.0069$^*$ &       -111.655$^*$ &       0.198$^*$ &        0.0930$^*$ &       -99.780$^*$ \\         
55864.3935     &        -8.4295     &      0.0077     &        -24.316     &       0.121     &        0.0082     &        -0.307 \\             
55933.2411     &        -1.2729     &      0.0097     &        -33.669     &       0.156     &        0.0215     &         0.368 \\             
56148.5790     &        -3.2947     &      0.0062     &       -31.1841     &      0.0985     &        0.0051     &        0.0375 \\             
56243.3345     &        10.5423     &      0.0068     &        -50.504     &       0.119     &        0.0156     &         0.127 \\             
56323.2640     &       -23.1077     &      0.0071     &         -3.531     &       0.114     &        0.0039     &        -0.121 \\             
56525.5388     &        -6.8886     &      0.0066     &        -25.907     &       0.110     &        0.0076     &         0.266 \\             
56526.5927     &        -8.4490     &      0.0067     &        -24.296     &       0.103     &       -0.0009     &        -0.301 \\             
56618.4321     &        10.9971     &      0.0073     &        -51.473     &       0.131     &        0.0001     &        -0.182 \\             
56889.5888     &         9.1280     &      0.0071     &        -48.721     &       0.120     &       -0.0035     &        -0.049 \\             
57414.2895     &       -22.0863     &      0.0102     &         -4.597     &       0.156     &       -0.0121     &         0.269 \\             
57664.4175     &       -27.4393     &      0.0070     &          2.693     &       0.111     &       -0.0103     &         0.042 \\             
57668.6030     &       -32.5635     &      0.0069     &          9.693     &       0.120     &       -0.0061     &        -0.157 \\             
57967.5713     &        -2.6955     &      0.0067     &        -31.913     &       0.106     &       -0.0187     &         0.183 \\             
58049.5126     &       -38.0529     &      0.0070     &         17.613     &       0.120     &        0.0070     &         0.038 \\             
58104.3403     &       -38.0874     &      0.0075     &         17.640     &       0.129     &        0.0045     &         0.021 \\             
&&&&&&   \\                                                                                                             
&&&&&&   \\                                                                                                             
\end{tabular}                                                                                                           

\end{minipage}%
\begin{minipage}{89mm}
\begin{tabular}{@{}l@{~~}c@{~~}c@{~~}c@{~~}c@{~~}c@{~~}c@{}}                                                            
\hline                                                                                                                  
\multicolumn{7}{c}{HIP 20601} \\                                                                                        
&&&&&&   \\                                                                                                             
BJD      & $RV_1$        & $\sigma_{RV 1}$ & $RV_2$        & $\sigma_{RV 2}$ & $O_1-C_1$   & $O_2-C_2$ \\               
-2400000     & km s$^{-1}$   & km s$^{-1}$      & km s$^{-1}$   & km s$^{-1}$      & km s$^{-1}$ & km s$^{-1}$ \\       
\hline                                                                                                                  
55532.4785     &        25.7705     &      0.0230     &         63.225     &       0.138     &       -0.0101     &         0.156 \\             
55965.3794$^*$ &        42.9021$^*$ &      0.0117$^*$ &        38.7895$^*$ &      0.0621$^*$ &       -0.0277$^*$ &       -1.1649$^*$ \\         
56243.5140     &        46.9362     &      0.0114     &        34.3063     &      0.0689     &        0.0042     &       -0.2538 \\             
56323.2404     &       -24.3260     &      0.0115     &       130.6220     &      0.0686     &       -0.0038     &        0.0230 \\             
56323.3136     &       -24.7308     &      0.0112     &       131.0056     &      0.0656     &        0.0106     &       -0.1584 \\             
56323.3628     &       -24.9478     &      0.0114     &       131.4262     &      0.0680     &        0.0058     &       -0.0239 \\             
56323.4538     &       -25.1850     &      0.0128     &       131.7206     &      0.0742     &       -0.0028     &       -0.0375 \\             
56323.5101     &       -25.2270     &      0.0147     &       131.7337     &      0.0825     &       -0.0191     &       -0.0592 \\             
56324.2438     &       -16.4954     &      0.0113     &       119.9919     &      0.0675     &       -0.0059     &       -0.0500 \\             
56324.4318     &       -12.1233     &      0.0123     &       114.2045     &      0.0712     &       -0.0008     &        0.0487 \\             
56324.4718     &       -11.1305     &      0.0128     &       113.0198     &      0.0742     &        0.0114     &        0.1856 \\             
56619.5265     &        33.7580     &      0.0139     &        52.5084     &      0.0877     &       -0.0049     &        0.1986 \\             
57009.4242     &        48.1266     &      0.0107     &        32.8635     &      0.0644     &        0.0088     &       -0.0982 \\             
57295.6072     &        49.3997     &      0.0117     &        31.3802     &      0.0681     &       -0.0030     &        0.1503 \\             
57729.4798     &       -10.4057     &      0.0084     &       111.8797     &      0.0533     &        0.0002     &        0.0376 \\             
57734.4367     &        30.0813     &      0.0093     &        57.3817     &      0.0587     &        0.0010     &        0.1084 \\             
57744.4419     &        47.8803     &      0.0104     &        33.1556     &      0.0577     &       -0.0076     &       -0.1161 \\             
&&&&&&   \\                                                                                                             
&&&&&&   \\                                                                                                             
\end{tabular}                                                                                                           

\end{minipage}\\%
\begin{minipage}{89mm}
\begin{tabular}{@{}l@{~~}c@{~~}c@{~~}c@{~~}c@{~~}c@{~~}c@{}}                                                            
\hline                                                                                                                  
\multicolumn{7}{c}{HIP 61732      } \\                                                                                  
&&&&&&   \\                                                                                                             
BJD      & $RV_1$        & $\sigma_{RV 1}$ & $RV_2$        & $\sigma_{RV 2}$ & $O_1-C_1$   & $O_2-C_2$ \\               
-2400000     & km s$^{-1}$   & km s$^{-1}$      & km s$^{-1}$   & km s$^{-1}$      & km s$^{-1}$ & km s$^{-1}$ \\       
\hline                                                                                                                  
55306.4207     &        -8.3478     &      0.0145     &       -26.6353     &      0.0463     &       -0.0958     &        0.2674 \\             
55605.5921     &       -23.0190     &      0.0224     &        -5.6286     &      0.0735     &       -0.0384     &        0.3460 \\             
55933.6824     &        -6.3513     &      0.0785     &        -29.404     &       0.238     &       -0.0083     &         0.211 \\             
55965.6664     &        -5.4211     &      0.0213     &       -30.4971     &      0.0680     &        0.0130     &        0.4097 \\             
56324.4596$^*$ &         5.5374$^*$ &      0.0495$^*$ &         12.427$^*$ &       0.132$^*$ &       23.7667$^*$ &        25.153$^*$ \\         
56700.6136     &       -22.9432     &      0.0199     &        -5.6340     &      0.0631     &       -0.0161     &        0.4167 \\             
56764.4370     &       -23.6411     &      0.0208     &        -4.7600     &      0.0680     &        0.0085     &        0.2643 \\             
57073.5168     &        -9.4719     &      0.0179     &       -24.8731     &      0.0589     &       -0.0082     &        0.3079 \\             
57159.4472     &        -5.4137     &      0.0193     &       -30.4919     &      0.0616     &        0.0451     &        0.3797 \\             
57786.6697     &        -7.6073     &      0.0315     &       -27.2952     &      0.0966     &        0.0131     &        0.5049 \\             
57790.6406     &        -8.1229     &      0.0251     &       -26.6537     &      0.0787     &        0.0416     &        0.3734 \\             
57882.3660     &       -22.3616     &      0.0218     &        -6.3860     &      0.0710     &        0.0608     &        0.3820 \\             
&&&&&&   \\                                                                                                             
&&&&&&   \\                                                                                                             
&&&&&&   \\                                                                                                             
\end{tabular}                                                                                                           

\end{minipage}%
\begin{minipage}{89mm}
\begin{tabular}{@{}l@{~~}c@{~~}c@{~~}c@{~~}c@{~~}c@{~~}c@{}}                                                            
\hline                                                                                                                  
\multicolumn{7}{c}{HIP 73449      } \\                                                                                  
&&&&&&   \\                                                                                                             
BJD      & $RV_1$        & $\sigma_{RV 1}$ & $RV_2$        & $\sigma_{RV 2}$ & $O_1-C_1$   & $O_2-C_2$ \\               
-2400000     & km s$^{-1}$   & km s$^{-1}$      & km s$^{-1}$   & km s$^{-1}$      & km s$^{-1}$ & km s$^{-1}$ \\       
\hline                                                                                                                  
55692.4882     &        -0.7934     &      0.0309     &        17.0707     &      0.0370     &       -0.0356     &       -0.0534 \\             
55784.3889     &         0.4684     &      0.0133     &        15.8849     &      0.0161     &       -0.0086     &        0.0049 \\             
56033.5122     &         4.2001     &      0.0141     &        12.5914     &      0.0162     &        0.2753     &        0.1855 \\             
56324.5998$^*$ &         8.1433$^*$ &      0.0150$^*$ &         8.1301$^*$ &      0.0211$^*$ &        0.4223$^*$ &       -0.4506$^*$ \\         
56414.4664$^*$ &         8.1472$^*$ &      0.0140$^*$ &         8.1379$^*$ &      0.0198$^*$ &       -0.6900$^*$ &        0.6818$^*$ \\         
56764.5111     &        12.9019     &      0.0135     &         3.1792     &      0.0159     &       -0.0959     &       -0.0845 \\             
57073.6441     &        16.3382     &      0.0149     &        -0.0219     &      0.0123     &        0.0428     &        0.0372 \\             
57159.4899     &        16.9948     &      0.0128     &        -0.7958     &      0.0154     &       -0.0239     &       -0.0080 \\             
57505.5554     &        15.5184     &      0.0127     &         0.6945     &      0.0154     &       -0.0165     &       -0.0127 \\             
57786.7050     &         1.5771     &      0.0169     &        14.7238     &      0.0204     &       -0.0281     &       -0.0194 \\             
57819.6003     &         0.2040     &      0.0124     &        16.1244     &      0.0151     &       -0.0205     &       -0.0099 \\             
57884.3865     &        -1.6977     &      0.0128     &        17.9910     &      0.0156     &       -0.0519     &       -0.0280 \\             
57907.4966     &        -2.0944     &      0.0128     &        18.4137     &      0.0155     &       -0.0384     &       -0.0186 \\             
&&&&&&   \\                                                                                                             
&&&&&&   \\                                                                                                             
\end{tabular}                                                                                                           

\end{minipage}\\%
\begin{minipage}{89mm}
\begin{tabular}{@{}l@{~~}c@{~~}c@{~~}c@{~~}c@{~~}c@{~~}c@{}}                                                            
\hline                                                                                                                  
\multicolumn{7}{c}{HIP 76006      } \\                                                                                  
&&&&&&   \\                                                                                                             
BJD      & $RV_1$        & $\sigma_{RV 1}$ & $RV_2$        & $\sigma_{RV 2}$ & $O_1-C_1$   & $O_2-C_2$ \\               
-2400000     & km s$^{-1}$   & km s$^{-1}$      & km s$^{-1}$   & km s$^{-1}$      & km s$^{-1}$ & km s$^{-1}$ \\       
\hline                                                                                                                  
55306.5029     &       -42.8153     &      0.0110     &       -52.7052     &      0.0222     &       -0.0892     &       -0.2235 \\             
55605.6601     &       -59.5963     &      0.0144     &       -32.1284     &      0.0343     &       -0.0035     &       -0.0147 \\             
55693.4977$^*$ &       -47.8939$^*$ &      0.0128$^*$ &       -55.4821$^*$ &      0.0361$^*$ &        2.3308$^*$ &      -12.0557$^*$ \\         
56033.5161     &       -39.9011     &      0.0141     &       -56.1313     &      0.0397     &       -0.0180     &       -0.2164 \\             
56148.3668     &       -67.1434     &      0.0144     &       -22.8255     &      0.0346     &        0.0040     &        0.1652 \\             
56324.5920$^*$ &       -47.6020$^*$ &      0.0144$^*$ &       -46.8863$^*$ &      0.0402$^*$ &        0.0042$^*$ &       -0.2977$^*$ \\         
56414.4732$^*$ &       -46.7105$^*$ &      0.0130$^*$ &       -39.0619$^*$ &      0.0372$^*$ &       -2.4072$^*$ &       11.5151$^*$ \\         
56526.3470     &       -41.2841     &      0.0134     &       -53.7254     &      0.0333     &        0.0682     &        0.4151 \\             
56763.6293     &       -60.6714     &      0.0151     &       -30.7782     &      0.0364     &        0.0160     &        0.0136 \\             
57073.6520     &       -42.0726     &      0.0108     &       -52.9364     &      0.0234     &        0.0988     &        0.2150 \\             
57159.4946     &       -40.3507     &      0.0130     &       -55.2787     &      0.0371     &       -0.0292     &        0.1067 \\             
57884.3934     &       -65.0381     &      0.0155     &       -25.4839     &      0.0367     &        0.0120     &        0.0396 \\             
57886.5146     &       -65.9159     &      0.0126     &       -24.3583     &      0.0302     &        0.0054     &        0.1130 \\             
57908.4167     &       -64.9054     &      0.0138     &       -25.6758     &      0.0330     &        0.0081     &        0.0125 \\             
57967.3659     &       -54.7292     &      0.0134     &       -38.7045     &      0.0381     &       -0.0725     &       -0.6301 \\             
&&&&&&   \\                                                                                                             
&&&&&&   \\                                                                                                             
\end{tabular}                                                                                                           

\end{minipage}%
\begin{minipage}{89mm}
\begin{tabular}{@{}l@{~~}c@{~~}c@{~~}c@{~~}c@{~~}c@{~~}c@{}}                                                            
\hline                                                                                                                  
\multicolumn{7}{c}{HIP 77725      } \\                                                                                  
&&&&&&   \\                                                                                                             
BJD      & $RV_1$        & $\sigma_{RV 1}$ & $RV_2$        & $\sigma_{RV 2}$ & $O_1-C_1$   & $O_2-C_2$ \\               
-2400000     & km s$^{-1}$   & km s$^{-1}$      & km s$^{-1}$   & km s$^{-1}$      & km s$^{-1}$ & km s$^{-1}$ \\       
\hline                                                                                                                  
56033.5344     &         5.4237     &      0.0469     &        -6.3398     &      0.0489     &        0.0317     &       -0.1054 \\             
56324.6204     &        -3.0082     &      0.0328     &         2.4322     &      0.0302     &        0.0543     &        0.1727 \\             
56413.5956     &        -4.1716     &      0.0305     &         3.4173     &      0.0279     &        0.0630     &       -0.0196 \\             
56414.4882     &        -4.1921     &      0.0300     &         3.4314     &      0.0270     &        0.0518     &       -0.0150 \\             
56525.3487     &        -5.1084     &      0.0291     &         4.1920     &      0.0326     &       -0.0903     &       -0.0322 \\             
56890.3385     &         6.2889     &      0.0341     &        -7.1557     &      0.0314     &        0.0230     &       -0.0434 \\             
57073.6786     &         4.2096     &      0.0231     &        -5.1364     &      0.0249     &       -0.1544     &        0.0652 \\             
57160.4459     &         1.4910     &      0.0265     &        -1.9056     &      0.0229     &        0.3734     &        0.0345 \\             
57505.5763     &        -4.9516     &      0.0320     &         3.9906     &      0.0348     &       -0.0967     &       -0.0695 \\             
57602.3763     &        -5.1568     &      0.0276     &         4.2329     &      0.0301     &       -0.0990     &       -0.0312 \\             
57883.4782     &         4.4851     &      0.0274     &        -5.4379     &      0.0294     &       -0.1273     &        0.0134 \\             
57915.4439     &         6.8188     &      0.0259     &        -7.7011     &      0.0281     &        0.0002     &       -0.0336 \\             
57967.3789     &         8.3892     &      0.0291     &        -9.2368     &      0.0264     &       -0.0259     &        0.0348 \\             
&&&&&&   \\                                                                                                             
&&&&&&   \\                                                                                                             
&&&&&&   \\                                                                                                             
&&&&&&   \\                                                                                                             
\end{tabular}                                                                                                           

\end{minipage}\\%
\vspace*{1cm}
\end{table*}

\addtocounter{table}{-1}
\begin{table*}
\caption{Continued.}
\scriptsize
\begin{minipage}{89mm}
\begin{tabular}{@{}l@{~~}c@{~~}c@{~~}c@{~~}c@{~~}c@{~~}c@{}}                                                            
\hline                                                                                                                  
\multicolumn{7}{c}{HIP 96656      } \\                                                                                  
&&&&&&   \\                                                                                                             
BJD      & $RV_1$        & $\sigma_{RV 1}$ & $RV_2$        & $\sigma_{RV 2}$ & $O_1-C_1$   & $O_2-C_2$ \\               
-2400000     & km s$^{-1}$   & km s$^{-1}$      & km s$^{-1}$   & km s$^{-1}$      & km s$^{-1}$ & km s$^{-1}$ \\       
\hline                                                                                                                  
54036.2377     &        -9.0509     &      0.0085     &         3.0741     &      0.0188     &       -0.0029     &       -0.0740 \\             
54242.5740     &        -9.8948     &      0.0073     &         4.0682     &      0.0157     &       -0.0029     &       -0.0055 \\             
54382.3061     &       -10.0356     &      0.0081     &         4.2243     &      0.0171     &       -0.0034     &       -0.0033 \\             
54609.5475     &        -9.7375     &      0.0077     &         3.9465     &      0.0172     &        0.0052     &        0.0364 \\             
56414.5879     &         0.7110     &      0.0074     &        -7.4156     &      0.0165     &        0.0174     &        0.1214 \\             
56525.3890     &         1.4861     &      0.0076     &        -8.3370     &      0.0172     &        0.0003     &        0.0689 \\             
56619.4076     &         2.1267     &      0.0084     &        -9.1430     &      0.0185     &       -0.0134     &       -0.0193 \\             
56890.4790     &         3.8014     &      0.0080     &       -11.0163     &      0.0173     &       -0.0001     &       -0.0705 \\             
57159.5549     &         4.6588     &      0.0079     &       -11.8947     &      0.0167     &       -0.0001     &       -0.0084 \\             
57295.3366     &         4.5014     &      0.0077     &       -11.7592     &      0.0166     &       -0.0042     &       -0.0409 \\             
57505.6141     &         3.1099     &      0.0124     &       -10.2134     &      0.0269     &        0.0125     &       -0.0398 \\             
57602.4588     &         1.9160     &      0.0081     &        -8.8603     &      0.0185     &       -0.0070     &        0.0252 \\             
58302.5008     &        -8.4244     &      0.0076     &         2.4612     &      0.0170     &        0.0085     &       -0.0123 \\             
58351.4943     &        -8.8211     &      0.0076     &         2.9015     &      0.0168     &       -0.0070     &        0.0100 \\             
&&&&&&   \\                                                                                                             
&&&&&&   \\                                                                                                             
\end{tabular}                                                                                                           

\end{minipage}%
\begin{minipage}{89mm}
\begin{tabular}{@{}l@{~~}c@{~~}c@{~~}c@{~~}c@{~~}c@{~~}c@{}}                                                            
\hline                                                                                                                  
\multicolumn{7}{c}{HIP 101452     } \\                                                                                  
&&&&&&   \\                                                                                                             
BJD      & $RV_1$        & $\sigma_{RV 1}$ & $RV_2$        & $\sigma_{RV 2}$ & $O_1-C_1$   & $O_2-C_2$ \\               
-2400000     & km s$^{-1}$   & km s$^{-1}$      & km s$^{-1}$   & km s$^{-1}$      & km s$^{-1}$ & km s$^{-1}$ \\       
\hline                                                                                                                  
55440.4261     &         16.939     &       0.211     &        -42.368     &       0.703     &         0.040     &         0.704 \\             
55784.4621$^*$ &         -6.452$^*$ &       0.147$^*$ &        -10.195$^*$ &       0.920$^*$ &        -0.838$^*$ &         0.273$^*$ \\         
56034.6277     &        -33.121     &       0.143     &         29.949     &       0.577     &         0.016     &         0.557 \\             
56147.4640     &         14.544     &       0.151     &        -40.641     &       0.519     &        -0.057     &        -0.898 \\             
56243.2835     &          9.778     &       0.152     &        -32.584     &       0.811     &        -0.000     &         0.176 \\             
57159.5702$^*$ &        -14.830$^*$ &       0.148$^*$ &          -1.19$^*$ &        1.05$^*$ &        -0.135$^*$ &         -3.88$^*$ \\         
57160.5648$^*$ &        -15.627$^*$ &       0.146$^*$ &         -1.383$^*$ &       0.683$^*$ &        -0.078$^*$ &        -5.302$^*$ \\         
57295.3811     &         9.8731     &      0.0925     &        -31.602     &       0.533     &        0.0335     &         1.247 \\             
57602.4958     &        -18.898     &       0.134     &          7.997     &       0.565     &        -0.063     &        -0.682 \\             
57633.4364     &         16.725     &       0.135     &        -43.591     &       0.599     &        -0.018     &        -0.744 \\             
57884.5573     &        -53.615     &       0.134     &         60.043     &       0.600     &         0.015     &         0.974 \\             
57967.4676     &        -41.757     &       0.134     &         41.743     &       0.717     &         0.090     &        -0.262 \\             
57969.4488     &        -47.624     &       0.139     &         50.173     &       0.579     &        -0.082     &        -0.079 \\             
58058.3569     &        -51.095     &       0.140     &         54.538     &       0.551     &        -0.005     &        -0.854 \\             
&&&&&&   \\                                                                                                             
&&&&&&   \\                                                                                                             
\end{tabular}                                                                                                           

\end{minipage}\\%
\begin{minipage}{89mm}
\begin{tabular}{@{}l@{~~}c@{~~}c@{~~}c@{~~}c@{~~}c@{~~}c@{}}                                                            
\hline                                                                                                                  
\multicolumn{7}{c}{HIP 104987     } \\                                                                                  
&&&&&&   \\                                                                                                             
BJD      & $RV_1$        & $\sigma_{RV 1}$ & $RV_2$        & $\sigma_{RV 2}$ & $O_1-C_1$   & $O_2-C_2$ \\               
-2400000     & km s$^{-1}$   & km s$^{-1}$      & km s$^{-1}$   & km s$^{-1}$      & km s$^{-1}$ & km s$^{-1}$ \\       
\hline                                                                                                                  
56889.4646     &       -26.5262     &      0.0065     &         -7.332     &       0.192     &       -0.0931     &        -3.174 \\             
56902.4309     &       -32.6324     &      0.0066     &          3.573     &       0.306     &       -0.0599     &         0.564 \\             
56904.5962     &       -32.6380     &      0.0068     &          3.440     &       0.262     &       -0.0374     &         0.397 \\             
56920.3467     &       -24.4103     &      0.0091     &         -7.056     &       0.298     &       -0.0155     &        -0.519 \\             
56923.3604     &       -21.5792     &      0.0065     &         -9.635     &       0.230     &       -0.0122     &         0.204 \\             
56935.2629     &        -9.7364     &      0.0066     &        -21.840     &       0.228     &       -0.0949     &         1.920 \\             
56964.2858     &        -4.2257     &      0.0065     &        -32.487     &       0.238     &       -0.0185     &        -2.383 \\             
56967.3061     &        -6.4232     &      0.0070     &        -26.063     &       0.208     &        0.0216     &         1.429 \\             
56967.3092     &        -6.4251     &      0.0071     &        -26.097     &       0.215     &        0.0221     &         1.393 \\             
57295.3939     &       -32.2195     &      0.0064     &          3.414     &       0.211     &       -0.0010     &         0.817 \\             
57305.3573     &       -31.2242     &      0.0072     &          2.262     &       0.505     &        0.0715     &         0.743 \\             
57602.5316     &       -30.8107     &      0.0064     &          1.103     &       0.438     &        0.1578     &        -0.034 \\             
57622.5357$^*$ &       -13.8302$^*$ &      0.0065$^*$ &        -18.162$^*$ &       0.258$^*$ &        0.0859$^*$ &         0.608$^*$ \\         
57634.4922     &        -3.6052     &      0.0065     &        -32.376     &       0.214     &        0.0110     &        -1.581 \\             
57967.5013$^*$ &       -17.9611$^*$ &      0.0064$^*$ &        -15.122$^*$ &       0.263$^*$ &        0.1778$^*$ &        -1.282$^*$ \\         
58041.2560     &        -0.1183     &      0.0063     &        -34.254     &       0.252     &        0.2328     &         0.352 \\             
&&&&&&   \\                                                                                                             
&&&&&&   \\                                                                                                             
\end{tabular}                                                                                                           

\end{minipage}%
\begin{minipage}{89mm}
\begin{tabular}{@{}l@{~~}c@{~~}c@{~~}c@{~~}c@{~~}c@{~~}c@{}}                                                            
\hline                                                                                                                  
\multicolumn{7}{c}{HIP 117186     } \\                                                                                  
&&&&&&   \\                                                                                                             
BJD      & $RV_1$        & $\sigma_{RV 1}$ & $RV_2$        & $\sigma_{RV 2}$ & $O_1-C_1$   & $O_2-C_2$ \\               
-2400000     & km s$^{-1}$   & km s$^{-1}$      & km s$^{-1}$   & km s$^{-1}$      & km s$^{-1}$ & km s$^{-1}$ \\       
\hline                                                                                                                  
55864.3650     &       -10.3379     &      0.0824     &       -35.4887     &      0.0450     &       -1.4658     &       -0.1402 \\             
56147.5270     &       -63.8632     &      0.0919     &        32.2578     &      0.0477     &       -0.8636     &       -0.1345 \\             
56243.3282     &       -31.7329     &      0.0865     &        -3.9741     &      0.0537     &        2.5450     &       -0.4209 \\             
56525.5154     &         0.3692     &      0.0969     &       -45.6023     &      0.0465     &        1.0037     &        0.0557 \\             
56619.4355     &         1.8052     &      0.0935     &       -47.2845     &      0.0460     &        1.1267     &        0.0166 \\             
56889.5626     &        -5.5758     &      0.0959     &       -40.7870     &      0.0533     &       -1.3766     &        0.4097 \\             
56948.4278     &        -4.2284     &      0.0907     &       -41.3813     &      0.0477     &       -0.2277     &        0.0637 \\             
57295.4350     &        -2.6845     &      0.0851     &       -44.2049     &      0.0434     &       -0.9645     &        0.0944 \\             
57349.4319     &       -62.3715     &      0.0769     &        31.5028     &      0.0470     &        0.0517     &       -0.1680 \\             
57352.3389     &       -55.3175     &      0.0961     &        22.9637     &      0.0478     &        0.0353     &        0.1415 \\             
57353.3302     &       -52.7521     &      0.0988     &        19.2277     &      0.0513     &       -0.4543     &        0.2287 \\             
57359.3294     &       -30.7757     &      0.0860     &        -4.1477     &      0.0468     &        2.6916     &        0.4199 \\             
57602.5586     &       -64.2228     &      0.0716     &        32.2981     &      0.0446     &       -1.2929     &       -0.0069 \\             
57967.5325$^*$ &       -17.3970$^*$ &      0.0902$^*$ &       -23.8904$^*$ &      0.0532$^*$ &       -0.7115$^*$ &        1.6796$^*$ \\         
58087.4119     &        -2.5781     &      0.0867     &       -44.7935     &      0.0471     &       -0.8063     &       -0.5589 \\             
&&&&&&   \\                                                                                                             
&&&&&&   \\                                                                                                             
&&&&&&   \\                                                                                                             
\end{tabular}                                                                                                           

\end{minipage}\\%
\vspace*{1cm}
\end{table*}

%===========================================================================================================================================

\section{Derivation of the spectroscopic orbits}
\label{sect:SB2orbits}

The SB2 orbits are calculated by fitting SB models with a Levenberg--Marquardt algorithm, thanks to the
routines in \cite{NumRec}.
However, it is necessary to operate in several steps in order to correct the uncertainties in the RVs.
The uncertainties of the RVs derived above are unreliable, which will lead to two types of error: 
first, the weights are inversely proportional to the squares of the uncertainties, and the RVs of
one component could be overweighted relative to those of the other. 
Second, even if the uncertainties lead to exact relative weights, the uncertainties inferred from
the covariance matrix will be false in the same proportions as the measurement uncertainties. 
For this reason, the uncertainties in Table~\ref{tab:RVs} are systematically corrected after each
orbit computation, by using the $F_2$ estimator of the goodness-of-fit defined in
\citet{Kendall}:

\begin{equation}\label{def:F2}
F_2= \left( \frac{9\nu}{2} \right)^{1/2} \left[ \left( \frac{\chi^2}{\nu} \right)^{1/3}+{\frac{2}{9 \nu}}-1 \right]
\end{equation}

\noindent
where $\nu$ is the number of degrees of freedom and $\chi^2$ is the weighted
sum of the squares of the differences between the predicted and the observed
values, normalized with respect to their uncertainties. When the predicted values
are obtained through a linear model, $F_2$ follows the normal distribution
${\cal N}(0,1)$. When non--linear models are used, but when the errors are small
in comparison to the measurements, as hereafter, the model is approximately linear around the
solution, and $F_2$ follows also ${\cal N}(0,1)$. 
Therefore, the $\chi^2$ of an orbit calculated with $\nu$ degrees of freedom must be close to 
the value corresponding to F2=0, which is:

\begin{equation}\label{def:khi20}
\chi^2_0 = \nu \left( 1 - \frac{2}{9 \nu} \right)^3
\end{equation}

The correction of the RV uncertainties is done as explained hereafter:
First, the components are treated separately. A noise $\varepsilon$ is added quadratically
to the RV uncertainties in order to have, for each component, an SB1 solution with $\chi^2 = \chi^2_0$. 
This gives the relative weights of the components
in the calculation of the SB2 solution. The final uncertainties are derived by applying a multiplying factor, $\varphi=\sqrt{\chi^2/\chi^2_0}$, 
so that the SB2 solution satisfies the condition $F_2=0$. Thus, the corrected uncertainties are given
by the equations~\ref{eq:correction1} and \ref{eq:correction2} hereafter:

\begin{align}
\sigma^\text{corr}_{RV, 1} &= \varphi_1 \times \sqrt{\sigma_{RV, 1}^2 + \varepsilon_1^2} \label{eq:correction1}\\
\sigma^\text{corr}_{RV, 2} &= \varphi_2 \times \sqrt{\sigma_{RV, 2}^2 + \varepsilon_2^2} \label{eq:correction2}
\end{align}

With the procedure described above, the coefficients $\varphi_1$ and $\varphi_2$ are necessarily equal.
However, it sometimes happens that the {\sc todmor} uncertainties of a component produce an SB1 orbit with a ${\chi}^2$
smaller than $\chi^2_0$. In this case, the noise $\varepsilon$ is null and the uncertainties are multiplied by the factor $\sqrt{\chi^2/\chi^2_0}$, which 
contributes to the $\varphi$--factor of the component. 

\begin{table*}
\caption{Correction terms applied to the uncertainties of the previous and of the new RV measurements. The composition of these terms into a uncertainty correction is set out in Section~\ref{sect:SB2orbits}, eqs.~\ref{eq:correction1} and~\ref{eq:correction2}. When the original publication provides only weights for the previous measurements,
$\varphi_{1,p}$ and $\varphi_{2,p}$ are the uncertainties corresponding to $W=1$, for the primary and for the secondary component, respectively.
}
\scriptsize
\begin{tabular}{cllcccccccc}
\hline
HIP & \#   & Reference of previous RV  &  \multicolumn{4}{c}{Correction terms for previous measurements} &  \multicolumn{4}{c}{Correction terms for new measurements}   \\
       &      &                 &$\varepsilon_{1,p}$&$\varphi_{1,p}$&$\varepsilon_{2,p}$&$\varphi_{2,p}$&$\varepsilon_{1,n}$&$\varphi_{1,n}$& $\varepsilon_{2,n}$& $\varphi_{2,n}$\\
       &      &                 &km s$^{-1}$        &               & km s$^{-1}$       &               & km s$^{-1}$       &               &  km s$^{-1}$       &         \\
\hline
HIP & 7134   & \cite{GrifEm75}  &       0           & 0.931         & $\ldots$          & $\ldots$      &  0.0123           & 1.009         & 0.1617             & 1.009   \\
HIP & 20601  & \cite{GrifGZG85}$^a$&  $\ldots$      &  $\ldots$     &  $\ldots$         &  $\ldots$     &  0                & 0.807         & 0.1052             & 1.029   \\ 
HIP & 61732  & \cite{HaMaUd12}$^a$ &  $\ldots$      &  $\ldots$     &  $\ldots$         &  $\ldots$     &  0.0357           & 1.292         & 0.                 & 1.292   \\ 
HIP & 73449  &\cite{Goldberg02}$^b$&    0           & 0.898         & 0                 & 0.714         &  0.0898           & 0.961         & 0.1183             & 0.961   \\ 
HIP & 76006  &\cite{Griffin05}$^c$ &    0           & 0.300         & 0                 & 1.147         &  0.0693           & 1.049         & 0.2621             & 1.049  \\ 
HIP & 77725  &\cite{Toko00}$^a$    &       $\ldots$ & $\ldots$      & $\ldots$          & $\ldots$      &  0.1277           & 1.116         & 0.0742             & 1.116   \\ 
HIP & 96656  & \cite{Balega07}$^a$ &  $\ldots$      &  $\ldots$     &  $\ldots$         &  $\ldots$     &  0.0060           & 1.255         & 0.0363             & 1.255   \\ 
HIP & 101452 & no RV published  &  $\ldots$         &  $\ldots$     &  $\ldots$         &  $\ldots$     &  0                & 0.491          & 0.8295             & 0.877  \\
HIP & 104987 &\cite{Massarotti08}&      0           & 0.605         & 0                 & 1.532         &  0.1167           & 0.857         & 2.0016             & 0.857  \\ 
HIP & 117186 &\cite{Nordstrom97} &      0           & 1.837         & 0                 & 1.487         &  1.0851           & 1.203         & 0.2761             & 1.203  \\ 
\hline
\label{tab:corsigRVprev}
\end{tabular}
\flushleft 
$^a$ The previous measurements don't improve the accuracy of the period, and they were not taken into account.\\
$^b$ The components were swapped before calculating the uncertainties.\\
$^c$ the radial velocities of the blend have been taken into account by assigning them an uncertainty of 0.340 km s$^{-1}$.
\end{table*}

The correction terms of the RV uncertainties of the 10 SB2 are listed in Table~\ref{tab:corsigRVprev}. HIP 20601 is the only exception
to the ``$\varphi_1=\varphi_2$'' rule, for the reason explained above.

The previously published RV measurements were also corrected. In most cases, the method was modified in order to maintain
the relative weights of the measurements. For this purpose, corrections are based solely on multiplicative coefficients $\varphi_1$
and $\varphi_2$. 
Since obtained through another process, these RVs are generally biased and they reduce the reliability of the other orbital parameters. 
Therefore, they are used only to recalculate the period: 
When the period from our measurements completed by previously published measurements is at least 4 times more accurate 
than that from our measurements alone, the orbital elements are derived from the SOPHIE RVs, setting the period to this new value.

\begin{table*}
 \centering
 \begin{minipage}{178mm}
  \caption{The orbital elements of the SB2s. The radial velocity of the barycentre, $V_0$, is in the reference 
system of the new measurements of the primary component. 
The minimum masses and minimum semi-major axes are derived from the true period 
($P_{true}=P \times (1-V_0/c)$). The numbers in parentheses refer to the previously published RV measurements that were taken into account,
in addition to the new ones, to derive the period, $P$; the other elements correspond to the new RVs alone.}
\scriptsize
  \begin{tabular}{@{}clrrrrrrrrrrrrrr@{}}
  \hline
HIP    & \#  & $P$          & $T_0$(BJD) & $e$          &  $V_0$     & $\omega_1$ & $K_1$      &${\cal M}_1 \sin^3 i$&$a_1 \sin i$&$N_1$& $d_{n-p}$   & $\sigma(O_1-C_1)$ \\
HD/BD  &     &              &            &              &            &            & $K_2$      &${\cal M}_2 \sin^3 i$&$a_2 \sin i$&$N_2$& $d_{2-1}$   &$\sigma(O_2-C_2)$    \\
       &     & (d)          & 2400000+   &              &(km s$^{-1}$)&($^{\rm o}$)&(km s$^{-1}$)&(${\cal M}_\odot$) &  (Gm)      &     &(km s$^{-1}$)&(km s$^{-1}$)    \\
  \hline
HIP & 7134   & 53.51164     & 55008.0873 & 0.39674      & -15.1227   & 279.717   & 24.6998     &0.5242     & 16.6843              &  16 &             &0.011   \\ 
HD  & 9313   &$\pm 0.00012$ &$\pm 0.0044$&$\pm 0.00027$ &$\pm 0.0038$&$\pm 0.037$&$\pm 0.0058$ &$\pm0.0025$&$\pm 0.0037$          &     &             &        \\
    &        &              &            &              &            &           & 34.674      &0.37342    & 23.422               &  16 & 0.4981      &0.201   \\
    &        &              &            &              &            &           &$\pm 0.075$  &$\pm0.00096$&$\pm 0.051$           &     &$\pm 0.0528$ &        \\
&&&&&&&&&&&\\ 
HIP & 20601  & 156.380540   & 56636.67055& 0.851280     & 41.5967    & 201.984   & 37.3352     &0.9050     & 42.1207              &  16 &            &0.0080 \\
HD  & 27935 &$\pm 0.000095$&$\pm 0.00046$&$\pm 0.000031$&$\pm 0.0041$&$\pm 0.015$&$\pm 0.0031$ &$\pm0.0016$&$\pm 0.0056$          &     &            &       \\
    &        &              &            &              &            &           & 50.322      &0.67143    & 56.772               &  16 & 0.1544     &0.131  \\ %
    &        &              &            &              &            &           &$\pm 0.040$  &$\pm0.00065$&$\pm 0.045$           &     &$\pm 0.0509$&        \\ %
&&&&&&&&&&&\\ 
HIP & 61732  & 595.18       & 57210.58   & 0.3393       & -15.956    & 64.85     & 9.197       &0.3326     & 70.81                &  11 &            &0.043  \\ % 
BD  &+17 2512&$\pm 0.20$    &$\pm 0.87$  &$\pm 0.0019$  &$\pm 0.032$ &$\pm 0.67$ &$\pm 0.021$  &$\pm0.0022$&$\pm 0.17$            &     &            &       \\ %
    &        &              &            &              &            &           & 13.068      &0.2341     &100.61                &  11 & 0.3466     &0.084  \\ %
    &        &              &            &              &            &           &$\pm 0.034$  &$\pm0.0014$&$\pm 0.27$            &     &$\pm 0.0736$&       \\ %
&&&&&&&&&&&\\ 
HIP &73449$^a$& 2528.6      & 57708.5    & 0.3752       & 8.147      & 99.54     & 10.175      &0.8928     & 327.9                &  11 & 0.3868     &0.097  \\ % 
HD  &132756  &$\pm 1.0$     &$\pm 4.0$   &$\pm 0.0023$  &$\pm 0.038$ &$\pm 0.74$ &$\pm 0.048$  &$\pm0.0080$&$\pm 1.5$             & (45)&$\pm 0.1434$&(0.862)\\ %
    &        &              &            &              &            &           & 10.253      &0.8861     & 330.4                &  11 & 0.0041     &0.069  \\ %
    &        &              &            &              &            &           &$\pm 0.037$  &$\pm0.0092$&$\pm 1.2$             & (45)&$\pm 0.0515$&(0.758)\\ 
&&&&&&&&&&&\\ 
HIP &76006$^a$&  581.816    & 56717.21   & 0.6460       & -47.331    &134.51     & 13.640      &0.4004     &  83.31               &  12 &-0.9549     &0.052  \\ % 
HD  &138525  &$\pm 0.025$   &$\pm 0.23$  &$\pm 0.0033$  &$\pm 0.031$ &$\pm 0.51$ &$\pm 0.048$  &$\pm0.0052$&$\pm 0.18$            & (51)&$\pm 0.0521$&(0.310)\\ %
    &        &              &            &              &            &           & 16.47       &0.3316     & 100.60               &  12 & 0.4104     &0.264  \\ %
    &        &              &            &              &            &           &$\pm 0.11$   &$\pm0.0027$&$\pm 0.61$            &(51,15$^b$)&$\pm 0.1039$&(1.317$^b$)\\%
&&&&&&&&&&&\\ 
HIP & 77725  & 1016.2       & 56911.2    & 0.3459       & -0.335     & 329.0     & 6.749       &0.1079     & 88.49                &  13 &            &0.135  \\ % 
BD  &+11 2874&$\pm 1.2$     &$\pm 2.3$   &$\pm 0.0063$  &$\pm 0.049$ &$\pm 1.3$  &$\pm 0.061$  &$\pm0.0016$&$\pm 0.75$            &     &            &       \\ %
    &        &              &            &              &            &           & 6.780       &0.1074     & 88.90                &  13 & -0.1453    &0.070  \\ %
    &        &              &            &              &            &           &$\pm 0.044$  &$\pm0.0020$&$\pm 0.50$            &     &$\pm 0.0722$&       \\ %
&&&&&&&&&&&\\ 
HIP & 96656  & 4350.0       & 57737.8    & 0.24294      & -3.2920    &  70.03    & 7.3500      &0.7879     &426.49                &  14 &            &0.008  \\ % 
HD  &186922  &$\pm 2.2$     &$\pm 2.7$   &$\pm 0.00066$ &$\pm 0.0061$&$\pm 0.26$ &$\pm 0.0049$ &$\pm0.0035$&$\pm 0.35$            &     &            &       \\ %
    &        &              &            &              &            &           & 8.062       &0.7183     &467.8                 &  14 &  0.1267    &0.053  \\ %
    &        &              &            &              &            &           &$\pm 0.017$  &$\pm0.0020$&$\pm 1.0$             &     &$\pm 0.0176$& \\ %
&&&&&&&&&&&\\ 
HIP & 101452 &87.6834       & 57097.926  & 0.6806       & -7.608     & 243.39    & 35.272      &1.361      & 31.158               &  11 &            &0.050  \\ % 
HD  & 196133 &$\pm 0.0020$  &$\pm 0.049$ &$\pm 0.0013$  &$\pm 0.058$ &$\pm 0.14$ &$\pm 0.045$  &$\pm0.020$ &$\pm 0.039$           &     &            &       \\ %
    &        &              &            &              &            &           & 51.08       &0.9396     & 45.12                &  11 &  0.0295    &0.772  \\ %
    &        &              &            &              &            &           &$\pm 0.34$   &$\pm0.0080$&$\pm 0.30$            &     &$\pm 0.3153$&       \\ %
&&&&&&&&&&&\\ 
HIP & 104987$^a$&98.8051    &57249.523   & 0$^c$        & -16.445    &  0$^c$    & 16.190      &0.239      & 21.998               &  14 &-0.2766    &0.088  \\ % 
HD  &202447/8&$\pm 0.0023$  &$\pm 0.042$ & (fixed)      &$\pm 0.027$ & (fixed)   &$\pm 0.036$  &$\pm0.016$ &$\pm 0.049$           &(108)&$\pm 0.0678$&(0.645) \\ %
    &        &              &            &              &            &           & 18.92       &0.2043     & 25.70                &  14 &  0.6472   &1.470  \\ %
    &        &              &            &              &            &           &$\pm 0.61$   &$\pm0.0071$&$\pm 0.82$            &(108)&$\pm 0.4715$&(1.503)\\ %
&&&&&&&&&&&\\ 
HIP & 117186$^a$&85.8244    & 56402.66   & 0.3339       & -21.03     & 176.78    & 32.59       &1.636      & 36.26                &  14 &-0.3657     &1.367  \\ % 
HD  &202447/8&$\pm 0.0013$  &$\pm 0.15$  &$\pm 0.0043$  &$\pm 0.39$  &$\pm 0.67$ &$\pm 0.47$   &$\pm0.026$ &$\pm 0.52$            & (19)&$\pm 0.5651$&(1.980) \\ %
    &        &              &            &              &            &           & 40.79       &1.307      & 45.38                &  14 & 0.9052     &0.276  \\ %
    &        &              &            &              &            &           &$\pm 0.14$   &$\pm0.037$ &$\pm 0.18$            & (19)&$\pm 0.4392$&(1.175) \\ %
 \hline
\label{tab:orbSB2}
\end{tabular}
\end{minipage}
\flushleft
$^a$ The elements were derived fixing $P$ to the value obtained taking also the previous measurements into account. \\
$^b$ Fifteen blend measurements were taken into account, with $\sigma(O-C)=0.342$ km s$^{-1}$. \\
$^c$ We have assumed a circular orbit since our RVs give the eccentricity $e=0.00000002 \pm 0.0022$.
\end{table*}

\begin{figure*}
\includegraphics[clip=,height=15.7 cm]{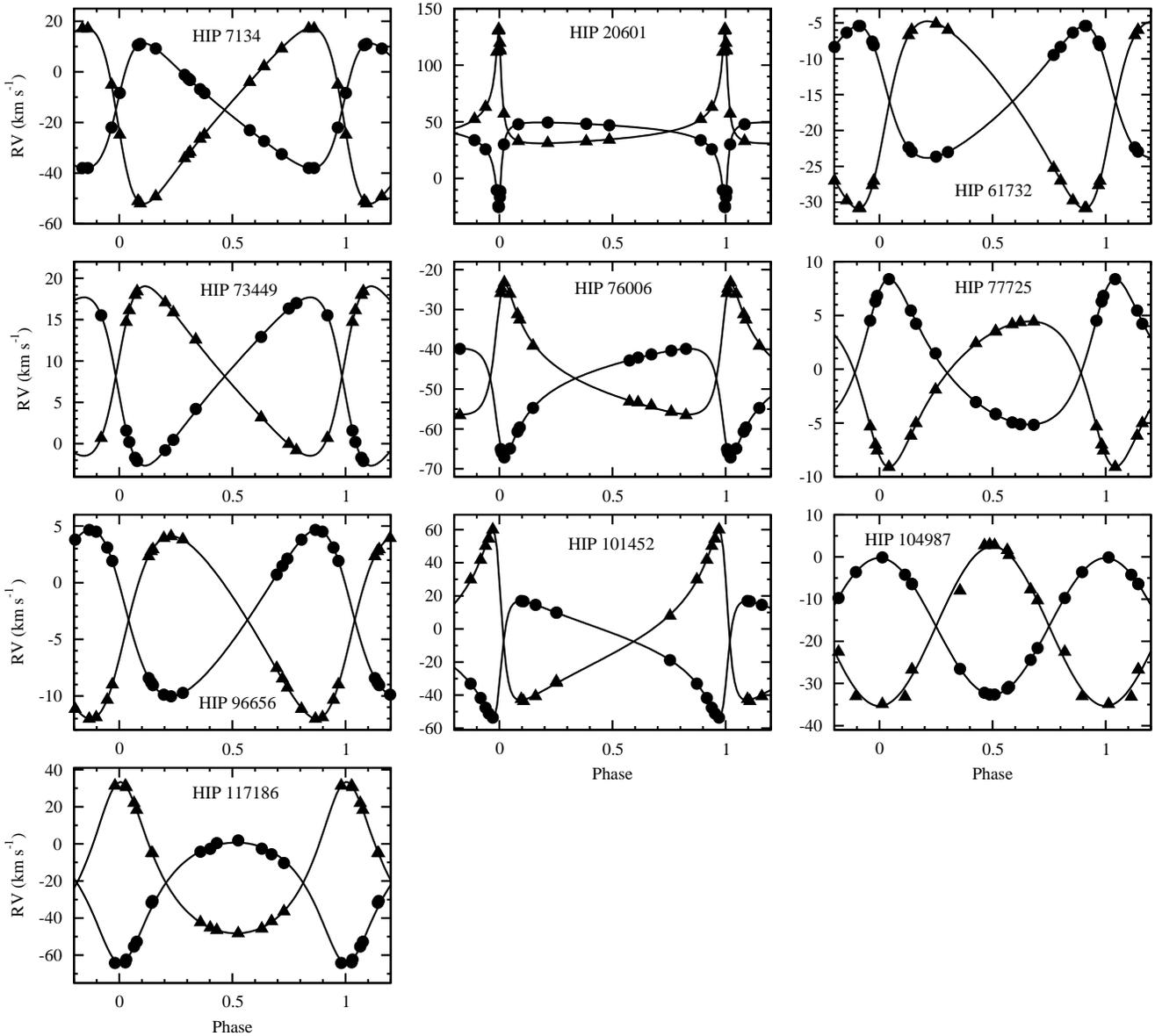} 
 \caption{The spectroscopic orbits of the 10 SB2; the circles refer to the primary component, and the
triangles to the secondary. For each SB2, the RVs are shifted to the
zero point of the SOPHIE measurements of the primary component.}
\label{fig:orbSB2}
\end{figure*}

\begin{figure*}
\includegraphics[clip=,height=15.7 cm]{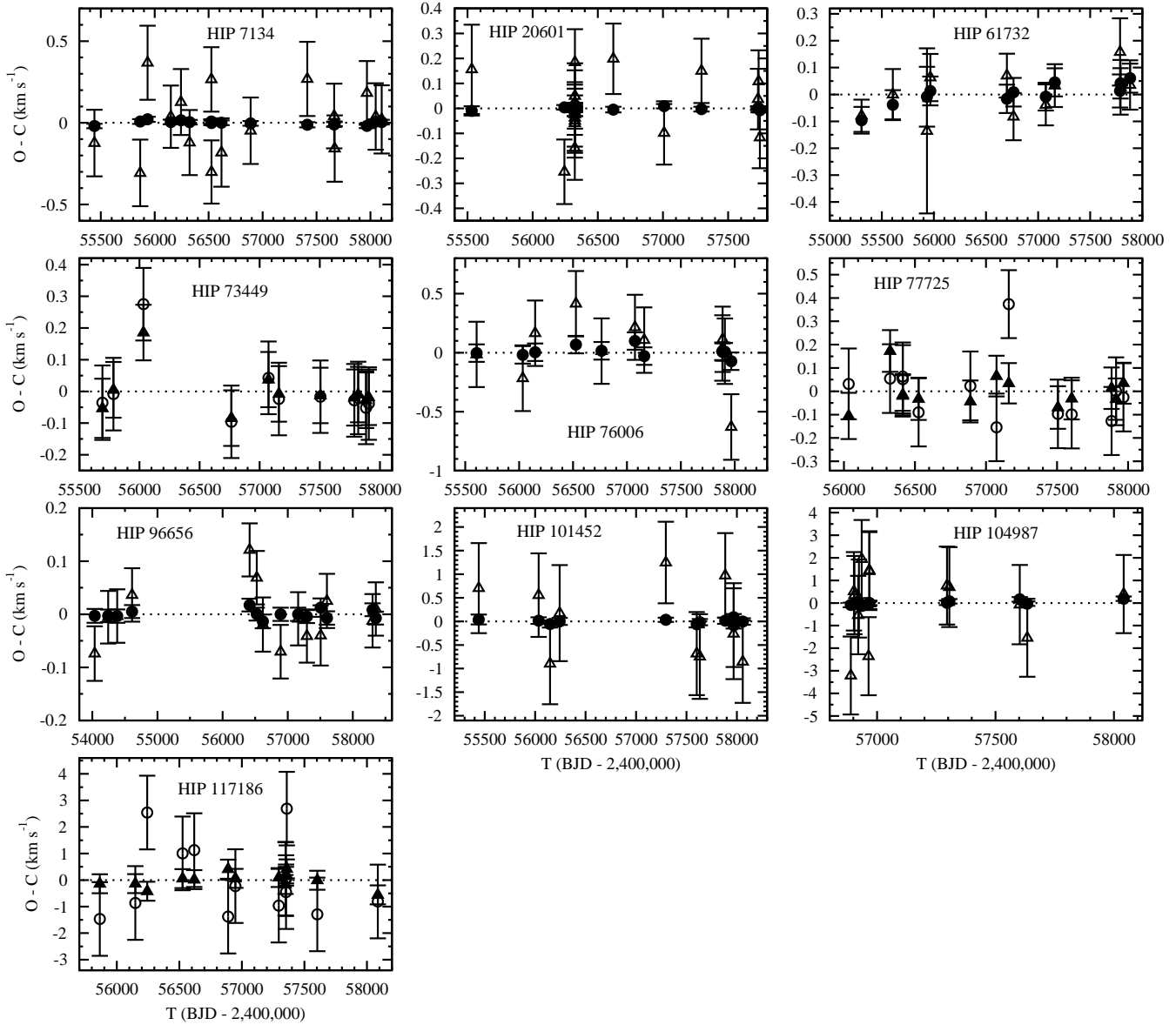} 
\caption{The residuals of the RVs obtained from {\sc todmor} for the 10 SB2s. The circles refer to the primary component, and the triangles to the secondary component. For readability, 
the residuals of the most accurate RV measurements are in filled symbols.}
\label{fig:resOrbSB2}
\end{figure*}

The orbital elements of the 10 stars are presented in Table~\ref{tab:orbSB2}. 
They include the following parameters: the period, $P$,  the periastron epoch, $T_0$, the eccentricity, $e$, the systemic radial
velocity, $V_0$, the periastron longitude, $\omega_1$, the RV semi--amplitudes of both components, $K_1$ and $K_2$, and the offset of the RVs
of the secondary component from that of the primary component, $d_{2-1}$. When the previously published RV measurements were taken into account to
derive the period, the offset of the SOPHIE RVs, $d_{n-p}$, is indicated. The table also includes the minimum masses, ${\cal M}_1 \sin^3 i$ and
${\cal M}_2 \sin^3 i$, and the minimum semi-major axes, $a_1 \sin i$ and $a_2 \sin i$, which are derived from the solution terms.
The spectroscopic orbits are shown in Figure~\ref{fig:orbSB2} and the residuals are in Figure~\ref{fig:resOrbSB2}.

Among the ten SB2s, HIP 20601 belongs to a multiple system, since \cite{GrifGZG85} found a faint distant visual component.
This star was observed by the Gaia satellite as Gaia~DR2~3283823383389256064 \citep{GaiaDR2}. Its separation relative to HIP 20601 was
7.09 arcsec, well above the diameter of the fibre of the SOPHIE spectrometer, which is 3 arcsec. Although an observation
of HIP~20601 by poor seeing observation could still be contaminated at such a distance, the low magnitude of this companion
($G$ = 12.97~mag i.e. 4 magnitudes fainter than HIP~20601) makes such contamination perfectly negligible, as evidenced by the
low residuals in the SB2 orbit, Figure~\ref{fig:resOrbSB2}.

%===========================================================================================================================================

%%%%%%%%%%%%%%%%%%%%%%%%%%%%%%%%%%%%%%%%%%%%%%%%%%%%%%%%%%%%%%%%%%%%%%%%%%%%%%%%%%%%%%%%%%%%%%%%%%%%%%%%%%%%%

\section{Masses and parallaxes of five stars resolved by interferometry}
\label{sect:interfero}
 
\subsection{Calculation method}
\label{sect:correctionBV}

Of the ten SB2s, five were sufficiently observed by interferometry to calculate a combined orbit giving the masses of the components and the trigonometric parallax of the system.
However, before combining RVs and interferometric measurements, uncertainties must first be corrected. This has been done for the RVs in Section~\ref{sect:SB2orbits}, but remains to be done
for interferometric measurements. The method is similar to that applied to RV uncertainties in Section~\ref{sect:SB2orbits}:
The visual orbit is calculated and the $\chi^2$ is considered. From this we deduce the corrective
coefficient that must be applied to the uncertainties for $F_2=0$. After this correction, the combined spectroscopic and interferometric orbit is derived from the RVs and from the
relative positions. The solutions terms are $P$, $T_0$, $e$, $V_0$, $\omega_1$, and $d_{2-1}$ as for the SB2 orbits, but also the position angle of the ascending node, $\Omega$, the inclination
of the orbital plane, $i$, the masses of the components, ${\cal M}_1$ and ${\cal M}_2$, and the trigonometric parallax of the binary star, $\varpi$. The apparent semi--major axis, $a$, is
finally derived from $P$, ${\cal M}_1$,  ${\cal M}_2$ and $\varpi$.

The uncertainties of the solution terms are extracted from the variance-covariance matrix of the Levenberg--Marquardt calculation,
but the uncertainty of $a$ is estimated by simulations, as explained hereafter: The solution terms that we have derived are used to
calculate the RVs or the relative position for each observation epoch. Simulated measurements are produced by adding, to these 
model values, errors generated according to the ${\cal N}(0,\sigma_0)$ distribution, where $\sigma_0$ is the measurement uncertainty
we finally obtained. A value of $a$ is then calculated from the set of simulated measurements. The uncertainty of $a$ is the
standard deviation of the values thus obtained.

The simulation program was also used to verify that the uncertainty correction process presented above and in Section~\ref{sect:SB2orbits} 
does not introduce an error that would add to the uncertainties estimated from the Levenberg--Marquardt calculation.
We have implemented the correction of the uncertainties of the simulated measurements by a multiplicative coefficient in order to have 
a zero $\chi^2$ after the calculation of the SB1 orbit of each component, then the SB2 orbit, as well as the interferometric orbit.
It thus appeared that the standard deviations of the solution terms calculated in the simulations were not affected by this modification,
and remained 
equal to the uncertainties we had found. Although the uncertainty correction implemented in the simulation is slightly simpler than that 
applied to real measurements, this shows the robustness of the correction process. The simulations also allowed us to verify the absence
of anomalies in the correlations between the different orbital parameters.

The results obtained for the five binaries are presented in
Table~\ref{tab:VB+SBorb}. In this table, the standard deviation of the astrometric residuals of the combined solution is followed by the standard deviation of the astrometric-only solution,
in parentheses. A comparison between these two terms shows that they are quite close, and such a similarity also appears if one compares the standard 
deviations of the radial velocities, $\sigma_{(o-c)\;RV}$, with the values in Table~\ref{tab:orbSB2}. This resemblance reflects the compatibility between
the astrometric and spectroscopic contributions of the combined solution.

\begin{table*}
 \caption{The combined VB+SB2 solutions. For consistency with the SB orbits and with the
forthcoming astrometric orbit, $\omega$ refer to the motion of the primary
component. Except for HIP 77725 and HIP 96656, $\Delta H$ comes from the flux ratios deduced from the PIONIER observations.
The standard deviation of the astrometric residuals $\sigma_{(o-c)\;VB}$ in parentheses refers to the astrometric-only solution.
}
\begin{tabular}{@{}lccccc@{}}
\hline
  &  HIP\,20601 & HIP\,77725 &  HIP\,96656 & HIP\,104987 & HIP\,117186 \\
\hline
$P$ (days)              &  156.380534 $\pm$ 0.000094 & 1015.53 $\pm$ 0.55    &   4345.3  $\pm$ 1.4     & 98.80450 $\pm$ 0.00035  &   85.8364 $\pm$ 0.0064   \\
$T_0$ (BJD-2400000)     & 56636.67052 $\pm$ 0.00046  & 56904.5 $\pm$ 1.6     &  57739.1  $\pm$ 2.0     &  57277.7 $\pm$ 1.7      & 56402.368 $\pm$ 0.094   \\
$e$                     &    0.851282 $\pm$ 0.000031 &  0.3415 $\pm$  0.0017 &  0.24280  $\pm$ 0.00065 & 0.00417  $\pm$ 0.00076  &  0.32778  $\pm$ 0.00073 \\
$V_0$ (km~s$^{-1}$)     &     41.5968 $\pm$ 0.0041   & -0.356  $\pm$ 0.043   &  -3.2884  $\pm$ 0.0053  & -16.458  $\pm$ 0.027    &  -21.11   $\pm$ 0.36    \\
$\omega_1$ ($^{\rm o}$) &     201.983 $\pm$ 0.015    & 325.51  $\pm$ 0.81    &   70.18   $\pm$ 0.20    &   102.9  $\pm$ 6.3      &  175.50   $\pm$ 0.34    \\
$\Omega$($^{\rm o}$; eq. 2000) & 340.513 $\pm$ 0.055 &  120.07 $\pm$ 0.50    &   292.78  $\pm$ 0.16    & 216.57   $\pm$ 0.16     &   16.942  $\pm$ 0.047   \\
$i$  ($^{\rm o}$)       &     103.133  $\pm$ 0.072   &   36.49 $\pm$  0.76   &    80.377 $\pm$ 0.097   &  151.52  $\pm$ 0.28     &  88.047   $\pm$ 0.043    \\
$a$ (mas)               & 11.338       $\pm$ 0.022   & 105.59  $\pm$ 0.73    &    189.38 $\pm$ 0.63    &  12.105  $\pm$ 0.013    & 4.677     $\pm$ 0.034  \\
${\cal M}_1$ (${\cal M}_\odot$) & 0.9798  $\pm$ 0.0019  & 0.510 $\pm$ 0.029  &  0.8216   $\pm$ 0.0037  &    2.20  $\pm$ 0.16     &    1.647  $\pm$ 0.022   \\
${\cal M}_2$ (${\cal M}_\odot$) & 0.72697 $\pm$ 0.00094 & 0.508 $\pm$ 0.029  &  0.7491   $\pm$ 0.0022  &   1.883  $\pm$ 0.083    &    1.316  $\pm$ 0.034   \\
$\varpi$ (mas)          &  16.703     $\pm$ 0.034    &    53.1  $\pm$  1.3   &   31.26   $\pm$ 0.11    &   18.11  $\pm$ 0.24     &     8.551 $\pm$ 0.080  \\
$H$ (mag)               &      7.209 $\pm$ 0.047     &    8.489 $\pm$ 0.010  &   5.980   $\pm$ 0.023   &   2.442  $\pm$ 0.196    &     6.252 $\pm$ 0.031 \\
$\Delta H$ (mag)        &     0.9990 $\pm$ 0.0158    & 0.06$^a$ $\pm$ 0.02$^a$ &  0.44  $\pm$ 0.24$^b$ &  2.1303  $\pm$ 0.0286   &    0.8914 $\pm$ 0.0074  \\
$d_{2-1}$ (km~s$^{-1}$) &  0.154      $\pm$ 0.050    &   -0.104 $\pm$  0.053 &  0.119   $\pm$ 0.016    &   0.676  $\pm$ 0.471    &    1.067  $\pm$ 0.374  \\
$\sigma_{(o-c)\;VB}$ (mas) &  0.031  (0.024)         &    3.95     (3.75)    &   2.06     (2.07)       &   0.652     (0.665)     & 0.0080      (0.0081)     \\
$\sigma_{(o-c)\;RV}$ (km~s$^{-1}$)  & 0.0081, 0.131  &        0.116, 0.124   &   0.012, 0.050          &       0.090, 1.488      &    1.521, 0.298        \\
\hline
\end{tabular}
\flushleft
$^a$ according to \cite{Horch17}.\\
$^b$ from the data of \cite{Balega07}.\\
\label{tab:VB+SBorb}
\end{table*}

%----------------------------------------------------------

\subsection{HIP 20601}
\label{sect:HIP20601}

The interferometric measurements we obtained for this star have been published in Table~1 of Paper II, where the uncertainties 
were corrected so that the visual orbit had $F_2=0$. 
Unlike Paper II, the spectroscopic part now consists only of our {\sc sophie} observations, with the reduction by {\sc todmor} seen above,
which ensures much more reliable results. The parameters of the combined orbit are in Table~\ref{tab:VB+SBorb}. 
The masses of the components are determined with a remarkable accuracy of 0.19\% and 0.13\% (about twice better than in Paper II).
The orbit remains visually very close to that shown in Figure~1 of Paper~II, and it is useless to reproduce it here.
The trigonometric parallax is 4.6 $\sigma$ larger than that found in the second {\it Gaia} DR (DR2), which is (17.32 $\pm$ 0.13)~mas. The
difference is probably due to the orbital motion, which was ignored in the reduction of the {\it Gaia} DR2.
In addition, we note that our uncertainty is 3.8 times smaller than that of {\it Gaia} DR2.

%----------------------------------------------------------

\subsection{HIP 77725}
\label{sect:HIP77725}

\begin{table}
 \caption{
The interferometric measurements of HIP 77725, taken from the INT4 catalogue
and adapted to our purpose. $\rho$ is the apparent separation and $\theta$ is the position
angle of the secondary component. $\sigma_a$ and $\sigma_b$ are the 
semi-major axis and the semi-minor axis of the ellipsoid error, respectively; they are derived as explained in the text. 
$\theta_a$ is the position angle of the major axis of the ellipsoid error.
The position angles are all given for the equinox of the observation epoch.
}
\begin{tabular}{@{}crrllr@{}}
\hline
$T$-2,400,000 & $\rho$    & $\theta$        & $\sigma_a$ & $\sigma_b$   & $\theta_a$     \\
(BJD)         & (mas)     & ($^{\rm o}$)    & (mas)      & (mas)        & ($^{\rm o}$) \\
\hline
49115.344 &  109.0 &   56.4 &   7.66 &   6.04 & 146.4 \\
49116.257 &  103.0 &   56.2 &   7.24 &   6.04 & 146.2 \\
50178.528 &  115.3 &   68.4 &   4.05 &   4.03 & 158.4 \\
50591.726 &  102.0 &  135.7 &   7.17 &   6.04 &  45.7 \\
52393.977 &  132.0 &  101.9 &   4.03 &   2.78 & 101.9 \\
53898.921 &   68.0 &  293.5 &   4.03 &   3.11 & 113.5 \\
54638.756 &  107.0 &  140.4 &   6.04 &   4.89 & 140.4 \\
56725.859 &   90.9 &  156.2 &   1.21 &  0.638 & 156.2 \\
57085.842 &   84.8 &    5.0 &  0.298 &  0.201 &  95.0 \\
57220.616 &  103.8 &   50.9 &   1.21 &   1.09 &  50.9 \\
57220.616 &  103.4 &   51.7 &  0.727 &  0.604 & 141.7 \\
\hline
\end{tabular}
\label{tab:HIP77725-interfero} 
\end{table}

This star is the visual binary BAG 7 and the Sixth Catalogue of Orbits of Visual Binary Stars\footnote{https://www.usno.navy.mil/USNO/astrometry/optical-IR-prod/wds/orb6} 
mentions a combined visual and spectroscopic orbit by \cite{Toko00}.

We found several interferometric measurements in the INT4 catalogue, and we selected the 11 of them that had measurement uncertainties.
A component inversion was corrected for the observation of 2008.4717, and we put the measurements in the same format
as the PIONIER measurements: times in years were converted to Julian days, and the uncertainties on $\theta$ and $\rho$ were converted into uncertainty 
ellipsoids aligned with the apparent separation, $\rho$.

A first calculation of the visual orbit then gives a solution with $F_2=4.82$. The uncertainties $\sigma_a$ and $\sigma_b$ were therefore corrected by multiplying
them by 2.016 to obtain a visual orbit of $F_2=0$.
The positions thus transformed and the final uncertainties are in Table~\ref{tab:HIP77725-interfero}.

 \begin{figure}
\includegraphics[clip=,height=5.4 in]{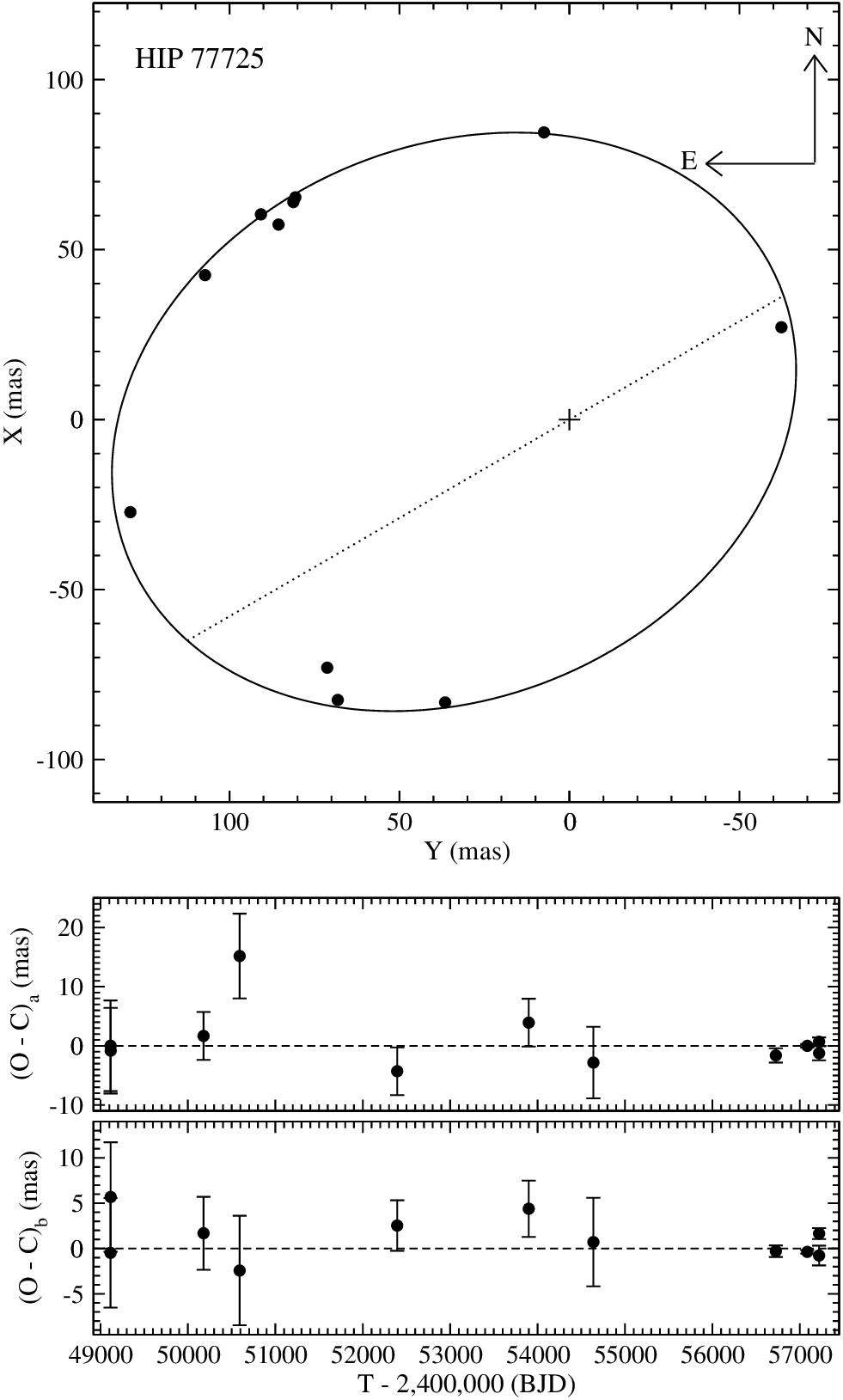}
 \caption{The visual part of the combined orbit of HIP 77725.Upper panel: the
visual orbit; the circles are the positions from Table~\ref{tab:HIP77725-interfero}; the node line is in dashes.
Middle panel: the residuals along the semi-major axis of the error ellipsoid. Lower panel: the residuals along the
semi-minor axis of the error ellipsoid.
}
\label{fig:HIP77725}
\end{figure}

The combined orbit was derived, and the solution terms in Table~\ref{tab:VB+SBorb} were obtained.
The interferometric orbit and the residuals are shown in Figure~\ref{fig:HIP77725}.
The masses of the components are slightly different but much more accurate than those derived by
\cite{Toko00}, which were ${\cal M}_1=(0.48 \pm 0.13)$~${\cal M}_\odot$ and ${\cal M}_2=(0.46 \pm 0.12)$~${\cal M}_\odot$.
The {\it Gaia} DR2 gives the trigonometric parallax $\varpi=47.29 \pm 0.17$~mas, which is 4.4 $\sigma$ smaller than
our result. Again, the difference may be due to the orbital motion.

%----------------------------------------------------------

\subsection{HIP 96656}
\label{sect:HIP96656}

This star is the nearby star GJ 765.2, and the double star MLR 224. 
\cite{Balega07} observed it with the 6m ``large altazimuth telescope'' (Russian: Bolshoi Teleskop Alt-azimutalnyi, or BTA6),
and obtained high-precision speckle measurements. They also took over visual measurements of lower quality, and combined
all these measurements with radial velocities measured with spectrovelocimeters. They thus determined the masses of the 
components with an accuracy of 2.4 to 2.5~\%. 

\begin{figure}
\includegraphics[clip=,height=4.94 in]{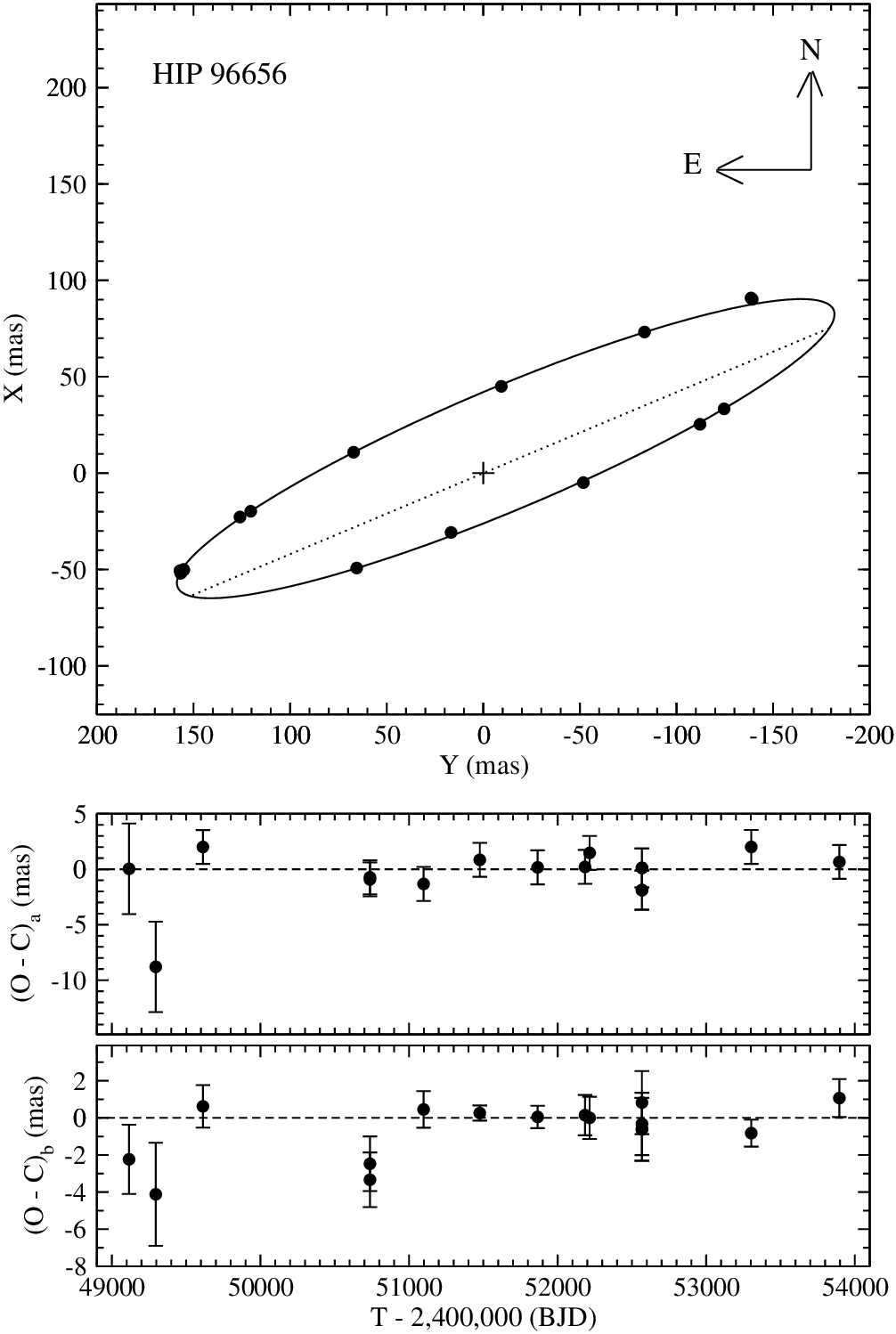}
 \caption{The visual part of the combined orbit of HIP 96656.Upper panel: the
visual orbit; the circles are the positions from Table~\ref{tab:HIP96656-interfero}; the node line is in dashes.
Middle panel: the residuals along the semi-major axis of the error ellipsoid. Lower panel: the residuals along the
semi-minor axis of the error ellipsoid.
}
\label{fig:HIP96656}
\end{figure}

We have ignored the visual measurements because of their poor quality, but have taken the BTA6 speckle measurements. 
As for HIP~77725, the epoch in years were converted in Julian days, and the parameters of the error ellipsoid,
$\sigma_a$, $\sigma_b$ and $\theta_a$, were derived from the position angle $\theta$ and from the uncertainties
on $\theta$ and $\rho$ given in Section 2 of the paper by Balega et al.;
the uncertainties were increased by 2.4~\% in order to obtain a visual orbit with $F_2=0$.
The measurement in Table~\ref{tab:HIP96656-interfero} were thus obtained.
By combining the speckle measurements and our radial velocities, we found a combined orbit including the
visual part presented in Figure~\ref{fig:HIP96656}. The elements are in Table~\ref{tab:VB+SBorb}.
The mass accuracy is now 0.45~\% for the primary component and 0.29~\% for the secondary, 5 and 8
times better than in Balega et al., respectively.

The parallax of the combined solution is $31.26 \pm 0.11$~mas, in disagreement with the $33.67 \pm 0.53$~mas given by
the {\it Gaia} DR2. Again, the difference probably comes from the orbital motion, neglected in the Gaia reduction.

\begin{table}
 \caption{
The interferometric measurements of HIP 96656, taken from Balega et al. (2007)
and adapted to our purpose. $\rho$ is the apparent separation and $\theta$ is the position
angle of the secondary component. $\sigma_a$ and $\sigma_b$ are the 
semi-major axis and the semi-minor axis of the ellipsoid error, respectively; they are derived as explained in the text. 
$\theta_a$ is the position angle of the major axis of the ellipsoid error.
The position angles are all given for the equinox of the observation epoch.
}
\begin{tabular}{@{}crrllr@{}}
\hline
 $T$-2,400,000 & $\rho$    &  $\theta$        &  $\sigma_a$ &  $\sigma_b$   &  $\theta_a$     \\
 (BJD)         &  (mas)     &  ($^{\rm o}$)    &  (mas)      &  (mas)        &  ($^{\rm o}$) \\
\hline
 49116.111 &    35.0 &   151.7 &    4.08 &    1.87 &  151.7 \\
 49296.029 &    52.0 &   264.7 &    4.08 &    2.78 &   84.7 \\
 49613.425 &   129.0 &   285.1 &    1.53 &    1.15 &  105.1 \\
 50736.983 &   165.7 &   303.3 &    1.53 &    1.47 &  123.3 \\
 50736.983 &   165.9 &   303.0 &    1.53 &    1.48 &  123.0 \\
 51097.842 &   111.0 &   311.3 &    1.53 &   0.988 &  131.3 \\
 51476.160 &    46.0 &   348.3 &    1.53 &   0.409 &  168.3 \\
 51865.253 &    68.0 &    80.8 &    1.53 &   0.605 &   80.8 \\
 52184.474 &   122.0 &    99.3 &    1.53 &    1.09 &   99.3 \\
 52215.483 &   128.0 &   100.2 &    1.53 &    1.14 &  100.2 \\
 52567.577 &   165.0 &   108.3 &    1.76 &    1.70 &  108.3 \\
 52567.577 &   165.0 &   107.8 &    1.76 &    1.70 &  107.8 \\
 52567.577 &   163.0 &   107.9 &    1.76 &    1.68 &  107.9 \\
 52567.577 &   163.0 &   107.8 &    1.76 &    1.68 &  107.8 \\
 53303.175 &    82.0 &   126.8 &    1.53 &   0.730 &  126.8 \\
 53895.963 &   115.0 &   282.6 &    1.53 &    1.02 &  102.6 \\
\hline
\end{tabular}
\label{tab:HIP96656-interfero}
\end{table}

%----------------------------------------------------------

\subsection{HIP 104987}
\label{sect:HIP104987}

\begin{table}
 \caption{
The interferometric measurements of HIP 104987.
$\rho$ is the apparent separation and $\theta$ is the position
angle of the secondary component, defined as the lightest.
$\sigma_a$ and $\sigma_b$ are the 
semi-major axis and the semi-minor axis of the ellipsoid error, respectively; they are 
corrected as explained in the text. $\theta_a$
is the position angle of the major axis of the ellipsoid error.
The position angles are given for the equinox of the observation epoch.
Observations dating back to before JD $2\,450\,000$ come from
Armstrong et al. (1992), after reversing the components. The others were carried out with the
Auxiliary Telescopes of the ESO VLTI, using the PIONIER instrument.
}
\begin{tabular}{@{}crrllr@{}}
\hline
$T$-2,400,000 & $\rho$    & $\theta$        & $\sigma_a$ & $\sigma_b$   & $\theta_a$     \\
(BJD)         & (mas)     & ($^{\rm o}$)    & (mas)      & (mas)        & ($^{\rm o}$) \\
\hline
47690.973 &   9.63 & 306.6 & 1.31   & 0.0540 &  90.9 \\
47695.903 &  10.30 & 283.0 & 0.303  & 0.0270 & 104.5 \\
47698.971 &  10.62 & 270.7 & 0.830  & 0.0607 &  90.8 \\
47700.944 &  10.38 & 262.3 & 2.15   & 0.108  &  94.7 \\
47715.955 &  12.39 & 213.8 & 2.59   & 0.385  &  80.6 \\
47720.922 &  11.79 & 198.3 & 2.49   & 1.51   & 114.8 \\
47746.781 &  10.25 &  96.7 & 1.69   & 0.324  & 106.5 \\
47747.841 &  10.76 &  92.2 & 0.938  & 0.243  &  88.9 \\
47751.822 &  11.33 &  78.0 & 0.668  & 0.121  &  88.6 \\
47758.762 &  11.55 &  53.5 & 0.303  & 0.148  &  78.1 \\
47761.720 &  11.62 &  41.5 & 0.762  & 0.148  & 108.2 \\
47765.738 &  12.36 &  36.0 & 1.93   & 0.0674 &  99.4 \\
47784.730 &  10.55 & 324.1 & 1.39   & 0.162  &  85.5 \\
47813.730 &  11.21 & 209.6 & 2.77   & 1.11   &  74.4 \\
47816.652 &  12.98 & 213.2 & 2.05   & 0.425  &  75.4 \\
47817.639 &  11.28 & 201.8 & 0.938  & 0.256  &  77.9 \\
47832.687 &  10.85 & 150.2 & 0.634  & 0.0607 &  69.9 \\
47836.668 &  10.66 & 134.5 & 0.492  & 0.0540 &  76.1 \\
48068.889 &  11.95 &  10.2 & 0.911  & 0.148  & 103.7 \\
48070.934 &  13.01 & 342.0 & 3.18   & 0.499  &  93.2 \\
48101.870 &  11.48 & 242.4 & 0.101  & 0.0270 &  86.4 \\
48102.856 &  11.73 & 239.7 & 0.223  & 0.0540 &  88.8 \\
48104.829 &  11.79 & 232.8 & 0.175  & 0.0338 &  94.9 \\
48130.834 &  10.68 & 143.4 & 0.128  & 0.0270 &  82.4 \\
48133.756 &  10.47 & 131.1 & 0.277  & 0.0472 &  86.3 \\
48134.778 &  10.52 & 127.1 & 0.196  & 0.0338 &  85.8 \\
48136.751 &  10.54 & 118.7 & 0.135  & 0.0338 &  96.2 \\
48137.810 &  11.41 & 112.4 & 0.445  & 0.0540 &  81.4 \\
48149.753 &  11.92 &  69.7 & 0.229  & 0.0540 &  81.4 \\
&&&&& \\
56937.598 &  11.20 &  89.85 & 0.0600 & 0.0248 & 132.0 \\
56939.599 &  11.37 &  82.22 & 0.0583 & 0.0229 & 135.0 \\
56948.578 &  12.06 &  51.41 & 0.0724 & 0.0212 & 137.0 \\
56949.537 &  12.09 &  48.49 & 0.0795 & 0.0530 & 147.0 \\
56962.538 &  11.69 &   5.35 & 0.0707 & 0.0300 & 134.0 \\
57537.870 &  11.73 &  63.44 & 0.0795 & 0.0406 & 172.0 \\
57569.873 &  10.63 & -50.42 & 0.1113 & 0.0883 & 119.0 \\
57597.710 &  12.02 &-151.36 & 0.0530 & 0.0300 &   4.0 \\
57599.740 &  11.96 &-158.26 & 0.0848 & 0.0335 &  15.0 \\
57600.725 &  11.88 &-161.20 & 0.1060 & 0.0459 & 147.0 \\
57622.707 &  10.77 & 115.44 & 0.0741 & 0.0371 & 152.0 \\
57625.659 &  10.88 & 103.52 & 0.0512 & 0.0318 &   3.0 \\
\hline
\end{tabular}
\label{tab:HIP104987-interfero}
\end{table}

\begin{figure}
\includegraphics[clip=,height=5.4 in]{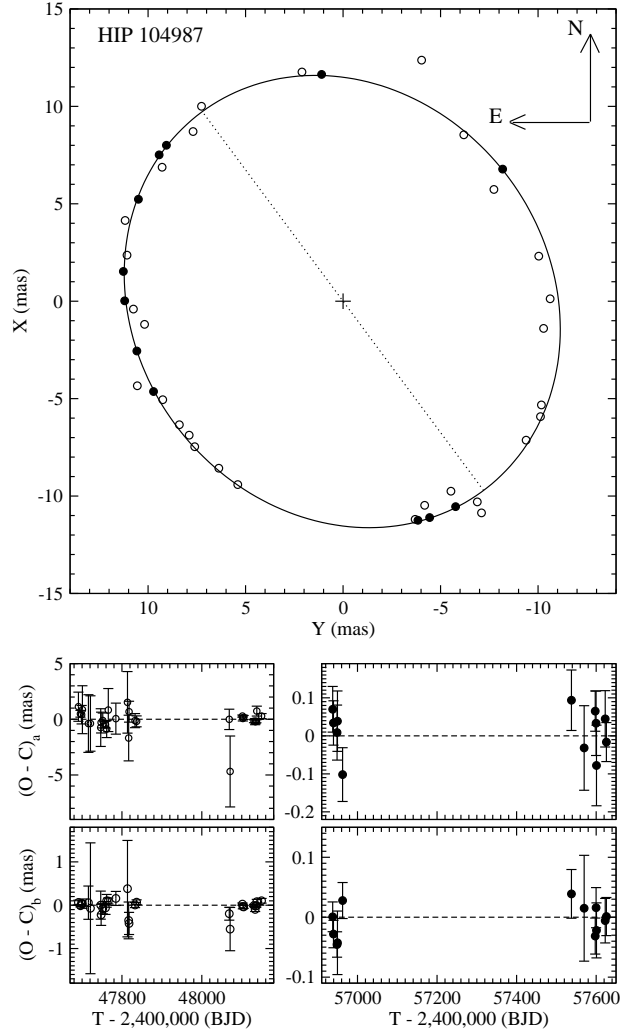}
 \caption{The visual part of the combined orbit of HIP 104987. Upper panel: the
visual orbit; the circles are the positions from Table~\ref{tab:HIP104987-interfero}; open circles refer to
the observations of Armstrong et al. (1992), while full circles represent our PIONIER measurements; the node line is in dashes.
Middle panels: the residuals along the semi-major axis of the error ellipsoid; the observations of Armstrong et al.
are in the left panel, and the PIONIER observations are in the right panel. Lower panels: the residuals along the
semi-minor axis of the error ellipsoid.
}
\label{fig:HIP104987}
\end{figure}

The INT4 catalogue contains many observations of this star, including a series of 29 measurements from the Mark III Optical
Interferometer published by \cite{MkT1992b}. 
These measurements are individually less accurate than PIONIER measurements, but their number still improves the accuracy of orbital parameters.  
We have therefore added them to our own measures, as follows:

\begin{itemize}

\item
The uncertainties of each set of measurements are corrected as explained in Section~\ref{sect:correctionBV}, by calculating
the interferometric orbit with each of them. A correction coefficient of 0.6086 is thus found for the uncertainties of the Armstrong's measurements,
and a coefficient of 0.1626 for ours.

\item
A comparison between the orbital elements from Armstrong et al. and those from PIONIER shows that the components have been inverted. 
Since PIONIER's position angles are compatible with the spectroscopic orbit, we correct the position angles of Armstrong et al. by 180 degrees. 

\item
The interferometric orbit is derived again from the two sets together. The $F_2$ estimator of the orbit is 1.03, inferring an acceptable compatibility
between the two sets. An additional correction of 1.0856 is still applied in order to have $F_2=0$.

\end{itemize}

The measurements with 
corrected uncertainties are presented in Table~\ref{tab:HIP104987-interfero}. They were used to derive the
combined spectroscopic and interferometric solution which is presented in Table~\ref{tab:VB+SBorb}, and 
in Figure~\ref{fig:HIP104987}.

The eccentricity of the orbit is very small for a binary with a period of nearly 100 days, and this is probably due to
the evolution of the primary component to the current G6 IV type. Our results provide a relevant
insight into the achievement of the circularization of the orbit: The SB2 solution given in Table~\ref{tab:orbSB2} is
circular since the calculation of an eccentric orbit lead to an eccentricity that is clearly not significant, with
$e/\sigma_e = 1.8\,10^{-8} \pm 2.24\,10^{-3}$; this is in agreement with \cite{Eggleton17}, who assumed that the
orbit is circular.
On the other hand, the eccentricity of the interferometric orbit is $(4.76 \pm 0.82)\,10^{-3}$.
This value is significant, compatible at the 2 $\sigma$-level  with the value of the SB2 solution, and we observe that the interferometric 
measurements from the INT4 catalogue give nearly the same periastron longitude as the PIONIER measurements: $\omega =(283.8 \pm 8.3)^{\rm o}$ for the
former and $\omega =(288 \pm 17)^{\rm o}$ for the latter. Therefore, we conclude that the orbit of HIP~104987 is not perfectly circular, and 
we adopt the eccentricity of the combined orbit, that is in Table~\ref{tab:VB+SBorb}; this value is significant at the 5.5 $\sigma$-level.

The masses we found are compatible with those of \cite{Eggleton17}, which are respectively 2.000 and 1.847 solar masses.

This star is not included in the {\it Gaia} DR2, but the parallax provided by the {\it Hipparcos 2} catalogue \citep{vanLeeuwen07}
is (17.14 $\pm$ 0.21) mas, which is 3.0 $\sigma$ less than our result. The difference may be due to an underestimation of the Hipparcos uncertainty.

%----------------------------------------------------------

\subsection{HIP 117186}
\label{sect:HIP117186}

As well as HIP 20601 and HIP 104987, this star was observed by interferometry with the PIONIER instrument.
Its measurements were presented in Paper II. By combining them with the RVs of this star, we have obtained 
the elements shown in Table~\ref{tab:VB+SBorb}.
These are not really different from the preliminary elements given in Paper II, but they are more reliable, 
due to the better quality and homogeneity of the RVs.

The parallax is 3.7 $\sigma$ larger than that found in the {\it Gaia} DR2, which is $(8.215 \pm 0.044)$~mas.

%----------------------------------------------------------

\subsection{Hertzsprung-Russell diagram and mass-luminosity relation}
\label{sect:diagHR-ML}

\begin{figure}
\includegraphics[clip=,height=2.7 in]{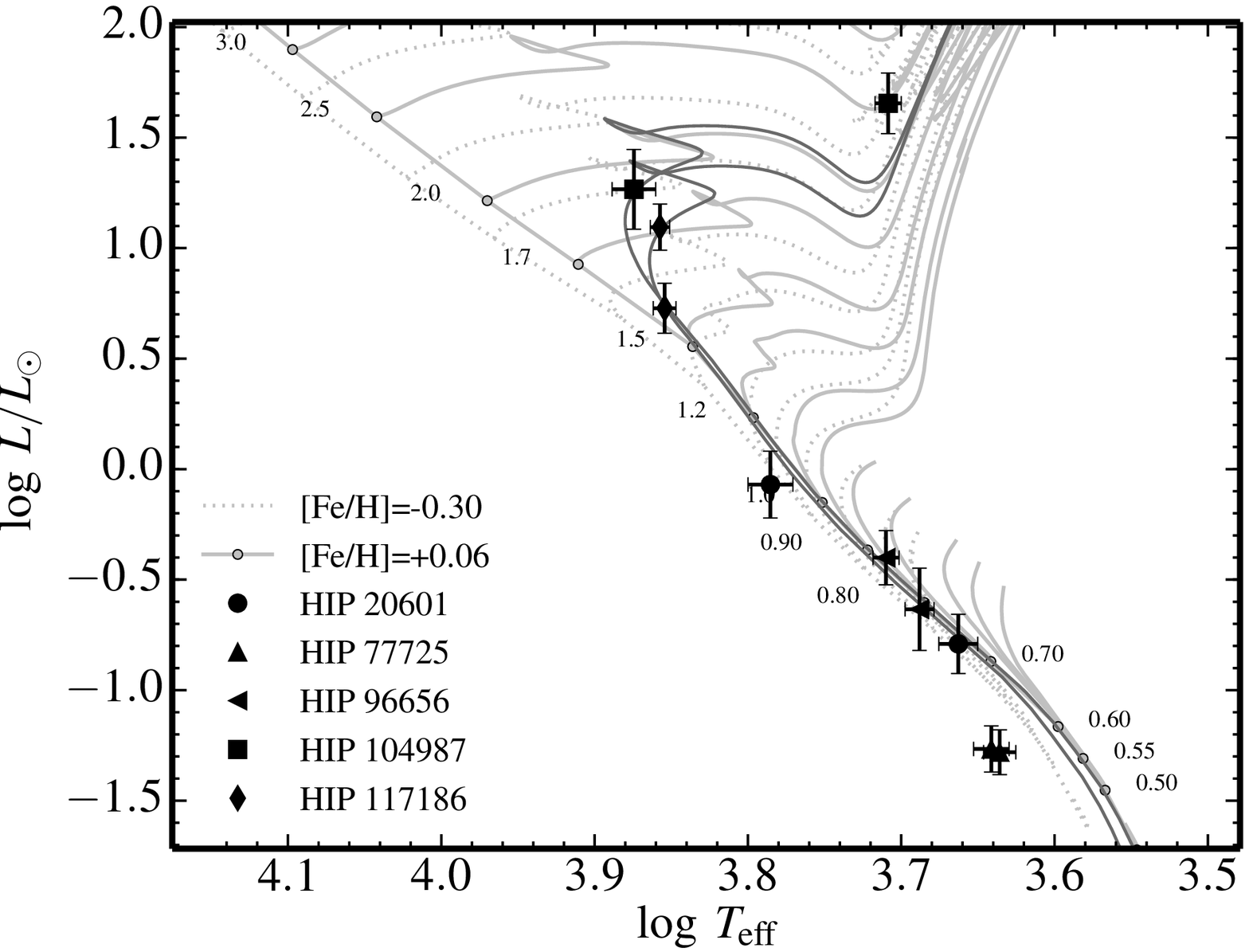}
 \caption{Hertzsprung-Russell diagram. The SB2 components are placed as well as stellar
evolutionary tracks and isochrones taken in the BaSTI database \citep{Hidalgo18}.
Evolutionary tracks in the mass range 0.5 to 3.0 ${\cal M}_\odot$ are plotted in light grey for two values of
the metallicity ([Fe/H] = $-0.30$ (dotted lines) and +0.06 (continuous lines)). In dark grey are
plotted from top to bottom isochrones of 1100 Myr at metallicity  [Fe/H]=$-0.03$ (HIP 104987) and 1500 Myr at metallicity [Fe/H]=$-0.11$ (HIP 117186).
}
\label{fig:diagHR}
\end{figure}

\begin{figure}
\includegraphics[clip=,height=2.7 in]{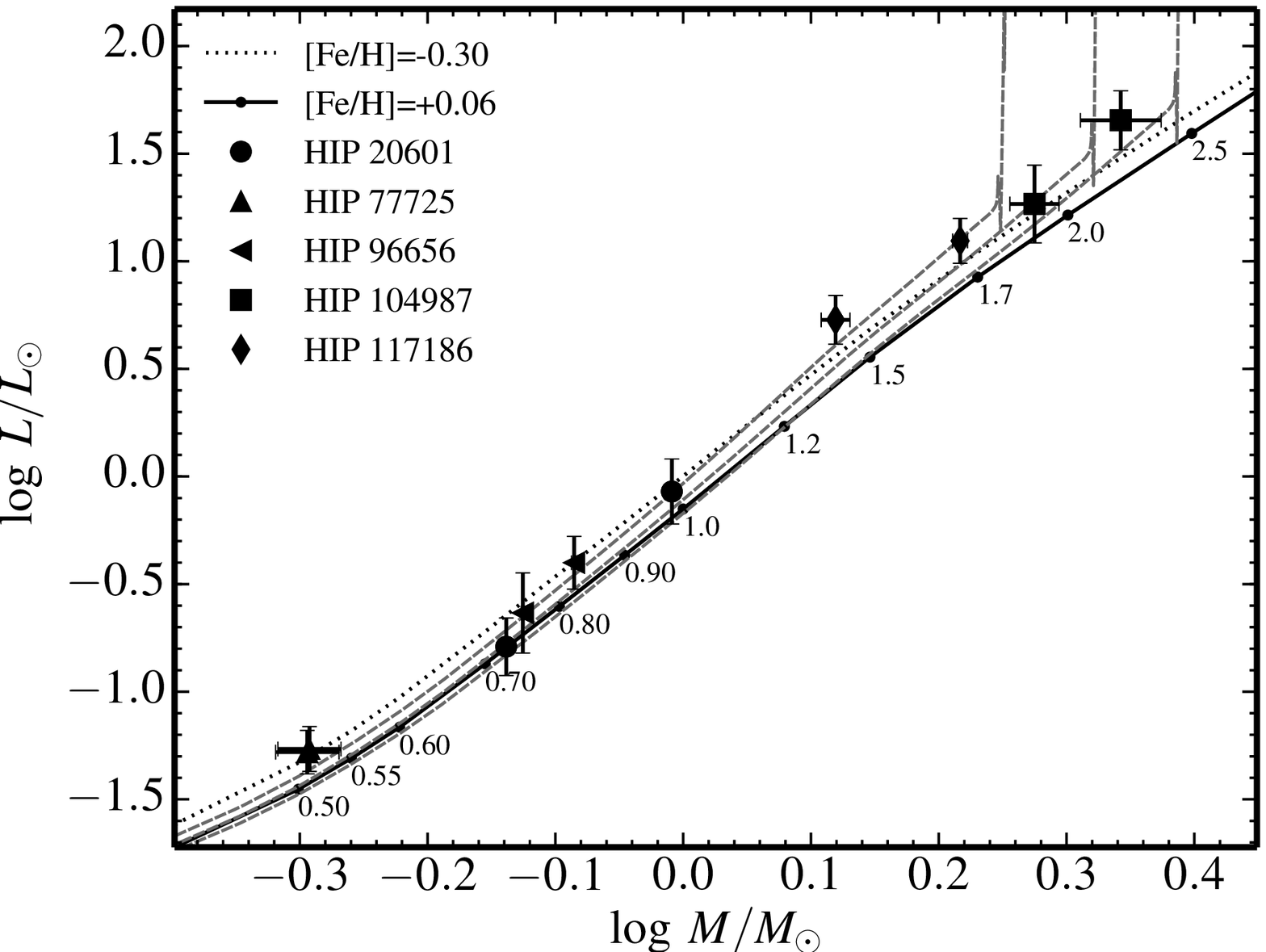}
 \caption{Mass-luminosity relation. The observed positions with error bars of the SB2 components are plotted. Black curves show the mass-luminosity relation on the zero age main sequence  at two values of the metallicity,  [Fe/H] = $-0.30$ (dotted line) and +0.06 (continuous
 line). Three isochrones are plotted with grey dashed lines. The isochrone at lower luminosities is for HIP 20601, that at higher luminosities  for HIP 117186 and in-between is the one for HIP 104987, see text for details.
}
\label{fig:ML}
\end{figure}

Figure~\ref{fig:diagHR} shows the location in the Hertzsprung-Russell (hereafter HR) diagram of the ten
stars in the five SB2 systems for which we have inferred the individual masses, while Figure~\ref{fig:ML} shows their position in the mass-luminosity plane.

To draw these figures, we started from the effective temperatures, metallicities, and surface
gravities given in Table~\ref{tab:stellpar}. We corrected the metallicity by adding $+0.12$ to each  [Fe/H]-value to account for the fact that in our study the solar spectrum is adjusted with  $\mathrm{[Fe/H]}_\odot = -0.12$ (as specified in the table caption). Following the conservative guidelines given in the footnotes of Table~\ref{tab:stellpar},  we added quadratically a systematic error of 100~K on the effective
temperature and 0.10 on [Fe/H].

For two systems, HIP 20601 and HIP 117186, the spectroscopic values in Table~\ref{tab:stellpar} differ from values in the literature. HIP 20601 is a probable member of the Hyades open cluster. Our corrected value of the metallicity ([Fe/H]=$-0.05 \pm 0.13$) differs by $2\sigma$ from the recent determination of the cluster average metallicity by \citet{Dutra-Ferreira16}, [Fe/H]=0.18$\pm$0.03 which is a robust value. If we fix the metallicity to this latter value and optimize again all other parameters, we get $T_\text{eff,A}$=6100$\pm$180\,K, $T_\text{eff,B}$=4600$\pm$90\,K, $\log g_A$=4.9$\pm$0.2, $\log g_B$=5.2$\pm$0.2, and unchanged $V\sin i$'s and flux ratio.
 As for HIP 117186, we noticed that the effective temperature of the A-component  ($T_\mathrm{eff, A}$ = 6208 $\pm$ 138~K) in Table~\ref{tab:stellpar} is smaller by more than 600~K than the value derived by \citet{Casagrande11} by photometric calibration ($T_\mathrm{eff, A}$ = 6853 $\pm$ 80 K),  while \citeauthor{Casagrande11}'s metallicity ([Fe/H]= $0.11\pm 0.10$) is higher than our corrected value, although still within the error bars. If we take  \citeauthor{Casagrande11}'s metallicity as a robust value and optimize again, we get $T_\text{eff,A}$=7200$\pm$30\,K, $T_\text{eff,B}$=7150$\pm$70\,K, $\log g_A$=3.8$\pm$0.2, $\log g_B$=4.1$\pm$0.2. In the following, we adopt these latter sets of parameters for the two couples.

To derive the individual luminosities, we proceeded in two steps. We first derived
the individual apparent magnitudes from the magnitude difference between the components
of HIP 20601, HIP 96656, HIP 104987, and HIP 117186 measured in the $H$-band by PIONIER. For HIP 77725, we used the magnitude difference in infrared measured by \citet{Horch17} with the
WYIN telescope. Then, we calculated the luminosities from the apparent magnitudes, the trigonometric
parallaxes of Table~\ref{tab:VB+SBorb}, and the bolometric corrections of \citet{Casagrande18},
the latter being functions of the effective temperature, metallicity, and surface gravity.

In Figures~\ref{fig:diagHR} and \ref{fig:ML}, we show the position of several stellar evolutionary tracks and isochrones
taken from the updated BaSTI database \citep[][]{Hidalgo18}. These models include state-of-the-art input physics and add 
overshooting of convective cores on the main sequence. As
indicated on the figure, the evolutionary tracks correspond to the range of metallicities and
masses of the studied SB2 members, while the isochrones fit the position of some of the stars.
We now discuss the figure for each SB2 couple:

\begin{itemize}
\item
HIP 20601. Due to their rather low mass, the stars are located in a region of the
HR diagram where the evolution proceeds quite slowly. Therefore, they cannot be age-dated in this diagram. Since the system is a member of the Hyades cluster, we can assume that like the Hyades, it is aged $\sim 600-700$ Myr \citep{Lebreton01}. Indeed, we can see in the mass-luminosity plane (Fig.~\ref{fig:ML}) that both components can be put on an isochrone of $700$ Myr corresponding to the Hyades metallicity.

\item
HIP 77725. The stars have very similar masses. Due to their low mass, they are located
in a region of the HR diagram where the evolution proceeds very slowly. Therefore their
age is mostly undetermined. We notice that while their effective temperatures as derived
in this study appear to be too hot with respect to those expected from the models, the
stars reasonably fit in the theoretical mass-luminosity relation.

\item 
HIP 96656. Again, the components have low mass and evolve slowly. In Fig.~\ref{fig:diagHR} the  position of the components are well-fitted on the zero age main sequence at their metallicity, however the system cannot
be age--dated accurately. Also, we note that both stars well fit in the theoretical mass-luminosity relation corresponding to their metallicity.

\item
HIP 104987. The two components can be positioned on isochrones of ages in the range
$\approx 1000 - 1200$~Myr (the isochrone plotted is of 1100~Myr). This age is older than the
age estimated by \citet{Griffin02} which is not surprising since both the observed
properties and stellar models have been considerably updated in the meantime. The
system is evolved: according to BaSTI stellar models, the A-component lies close to the base
of the red giant branch while
the B-one is reaching the end of the main sequence. Both components sit on their isochrone in the theoretical mass-luminosity plane.

\item
HIP 117186. First, we point out that if we had taken the spectroscopic values of the effective temperatures and metallicity in Table~\ref{tab:stellpar}, it would not have been possible to place the stars on the same isochrone. On the other hand, with the revised $T_\mathrm{eff}$ based on the metallicity value of \citet{Casagrande11}, the components sit on an isochrone of $1500$ Myr in the HR diagram as well as in the mass-luminosity plane.

\end{itemize}

We conclude that  the properties of the SB2 couples are overall well
retrieved by theoretical models in the mass-luminosity plane. Concerning the HR diagram, the fit is also satisfactory for the 5 systems, once the metallicity and, as a consequence, the effective temperatures of one of them (HIP 117186) have been revised after we adopted robust [Fe/H] determination from the literature. The masses we determined in this study are very precise, while
the characterization of the stars would benefit from further improvements in the determination
of their luminosities, effective temperatures, and metallicities. The error bars on these latter
are still too high to constrain the models. Furthermore, although very modern, stellar models
are still being affected by uncertainties in their input physics and initial helium abundance.
A full characterisation of the stars is well beyond the scope of the paper.

%________________________________________________________________________________________________________________________________

\section{Summary and conclusion}
\label{sect:conclusion}

We have obtained 146 spectra for 10 SB2s: 7 SB2s previoulsy known, and 3 binaries that were only SB1s. The RVs of the
components were derived with the {\sc todmor} cross-correlation algorithm. After discarding the RVs coming from 14 blended
spectra, we have derived the orbital elements of the 10 SB2s. We found minimal masses with an accuracy better than 1~\%
for 11 of the 20 components.

Five of the 10 SB2s have received enough long-baseline or speckle interferometric measurements to calculate the masses of their components. 
The PIONIER measurements of one of them, HIP 104987, were never published before. Thanks to these data, we have
derived the masses of the components of these five binary stars with accuracies ranging from 0.13\% to a few percents, and we have found that the orbit of HIP 104987 is not circular,
although its eccentricity is very small. We were also able to provide an estimate of the state of evolution of these stars
by placing them in the HR diagram.

Taking into account Paper III and IV, we have now accurate orbits for 34 SB2s, and combined SB2 and interferometric orbits for 9 of them.
These latter will be useful to check the masses that will be obtained from {\it Gaia} in the future, and they can also be
used to control the forthcoming DR3 parallaxes.

\section*{Acknowledgments}

This project was supported by the french INSU-CNRS ``Programme National de Physique Stellaire'', 
and the Centre National des Etudes Spatiales (CNES). We are grateful to the staff of the
Haute--Provence Observatory, and especially to Dr H. Le Coroller, Dr M. V\'{e}ron, and the night assistants, for their
kind assistance. 
{\it PIONIER} is funded by the Universit\'e Joseph Fourier (UJF), the Institut de Plan\'etologie et d'Astrophysique de Grenoble (IPAG),
and the Agence Nationale pour la Recherche (ANR-06-BLAN-0421, ANR-10-BLAN-0505, ANR-10-LABX56).
The integrated optics beam combiner is the result of a collaboration between IPAG and CEA-LETI based on CNES R\&T funding.
We made use of the SIMBAD database, operated at CDS, Strasbourg, France.

\label{lastpage}

\end{document}